\documentstyle[preprint,aps,prb,amsfonts,amssymb]{revtex}
\begin{document}
\draft
\title{ 
        Quasi-particle description for the transport through
\\
       a small interacting system 
}
\author{Akira Oguri}
\address{
          Department of Material Science,
          Faculty of Science,
          Osaka City University, \\
          Sumiyoshi-ku, Osaka 558-8585,
          Japan
}

\date{\today}

\maketitle

\begin{abstract}
We study effects of electron correlation on the transport through 
a small interacting system connected to reservoirs 
using an effective Hamiltonian which describes the free quasi-particles 
of a Fermi liquid. 
The effective Hamiltonian is defined microscopically 
with the value of the self-energy at $\omega=0$. 
Specifically, we apply the method to a Hubbard chain 
of finite size $N$ ($=1,\,2,\,3,\, \ldots$), 
and calculate the self-energy 
within the second order in $U$ in 
the electron-hole symmetric case.  
When the couplings between the chain and the reservoirs 
on the left and right are small, 
the conductance for even $N$ decreases with increasing $N$ 
showing a tendency toward a Mott-Hubbard insulator.
This is caused by the off-diagonal element of the self-energy, 
and this behavior is qualitatively different from that 
in the special case examined in the previous work.
We also study the effects of the asymmetry in the two couplings.
While the perfect transmission due to the Kondo resonance 
occurs for any odd $N$ in the symmetric coupling, 
the conductance for odd $N$ decreases with increasing $N$ in the case of
the asymmetric coupling. 
\end{abstract}

\bigskip
\bigskip

\pacs{PACS numbers: 72.10.-d, 72.10.Bg, 73.40.-c}


\narrowtext

\section{INTRODUCTION}
\label{sec:intro}

Effects of the inter-electron interaction on 
the transport through small interacting systems 
have been a subject of current interest. 
For instance, the Kondo effect expected to be observed 
in a quantum dot\cite{NL,GR,Kawabata} has been studied 
intensively from both theoretical\cite{MWL1-2,HDW2}
and experimental sides.\cite{RalphBuhrman,Goldharber-Gordon,Kouwenhoven} 
Also, the Tomonaga-Luttinger behavior expected in 
one-dimensional systems\cite{KaneFisher,FurusakiNagaosa} 
has been investigated in a quantum wire.\cite{Tarucha}

For studying the quantum transport of small systems theoretically, 
a formulation which is able to treat both the interaction 
and interference effects in a unified way is necessary.
Meir and Wingreen\cite{MW} have presented   
one such framework using the nonequilibrium Keldysh formalism.
\cite{Keldysh,Caroli}  
Especially, Eq.\ (6) of Ref.\ \onlinecite{MW} is 
a general formula for the total current, and is available for various systems.
While the quantum transport theory for nonequilibrium states 
is important to understand the physics in small systems, 
there are some ambiguities for determining the stationary state. 
So, it seems to be still meaningful to describe   
a formulation based on the linear response theory. 
In the Kubo formalism, 
the conductance for non-interacting electrons at $T=0$
can be expressed in terms of 
the Green's function at Fermi energy.\cite{Economou,Fisher,Lee} 
This is also true for interacting electrons 
if the ground state is a Fermi liquid.\cite{ao5,ao6} 
The purpose of this paper is to apply  
a quasi-particle description of a Fermi liquid to the conductance 
of a small interacting systems.
At $T=0$, an effective Hamiltonian for free quasi-particles
can be defined microscopically based on the perturbation theory,
which reproduces the value of the Green's function at $\omega=0$. 
This kind of an effective Hamiltonian has been introduced, for instance,
for the single Anderson impurity and systems with a translational
invariance.\cite{Hewson} 
One of the characteristics of the small interacting system is that
the self-energy has off-diagonal elements. This is because 
generally the system consists of a number of resonant states 
and has no translational invariance.

Specifically, we apply the method to a Hubbard chain of finite 
size $N$ ($=1,\,2,\,3,\, \ldots$) connected to non-interacting leads.
This system can be regarded as a model 
for a Mott-Hubbard insulator of nanometer scale. 
Also, for small $N$, 
the system can regarded as a model for a series of quantum dots, 
which has been examined by advanced numerical methods  
such as the numerical renormalization group\cite{Izumida1-3}
and quantum Monte Carlo methods.\cite{ao6,Sakai,ao} 
It may also be considered as a model for a quantum wire, 
which has been studied with the bosonization approach 
take into account the Umklapp scattering.
\cite{PonomarenkoNagaosa,OdintsovTokura,StarykhMaslov,MoriOgataFukuyama,FujimotoKawakami} 
We calculate all the elements of a $N \times N$ matrix self-energy 
within the second order in $U$ in the electron-hole symmetric case.  
The unperturbed Hamiltonian is taken to be connected 
including the coupling between the Hubbard chain and leads in it
as that in the case of the Anderson impurity.\cite{YY1-2}
In the previous paper,\cite{ao7}
we have reported the results obtained in a special case 
$v_L^{\phantom{\dagger}} = v_R^{\phantom{\dagger}} =t$,
where $v_L^{\phantom{\dagger}}$ ($v_R^{\phantom{\dagger}}$) 
is the tunneling matrix element between the chain and the left (right) lead,
and $t$ is the hopping matrix element of the chain.
In this case, one can obtain the self-energy analytically 
owing to the property that the non-interacting system has 
the translational invariance.
However, this particular feature is realized only in this case,
and some behaviors seeing in the results seem to depend on it: 
the reduction of the conductance 
for even $N$ is proportional to $U^4$ for small $U$,  
and in the limit of large even $N$ the conductance tends to be finite.

In the present work, 
we examine the dependence of the conductance on the coupling 
$v_L^{\phantom{\dagger}}$ and $v_R^{\phantom{\dagger}}$ calculating 
the self-energy numerically. 
In contrast to the special case mentioned above,   
the results obtained in weak 
coupling cases $v_{L}^{\phantom{0}}$, $v_{R}^{\phantom{0}}<t$ show 
quite different behaviors:
the reduction of the conductance 
for even $N$ is proportional to $U^2$ for small $U$, and  
the conductance tends to zero for large even $N$. 
Physically, 
the difference between the weak coupling 
cases $v_L^{\phantom{\dagger}}$, $v_R^{\phantom{\dagger}}<t$ and
the special case $v_L^{\phantom{\dagger}} 
= v_R^{\phantom{\dagger}} =t$ is caused by whether $N$ revels 
in the chain form distinguishable resonant states or not. 
The effect of the interaction is enhanced for the weak coupling cases.
It is also shown generally beyond the second-order perturbation 
that the perfect transmission occurs for odd $N$ when 
the system has both the inversion symmetry $v_L^{\phantom{\dagger}} 
= v_R^{\phantom{\dagger}}$ ($\equiv v$) and 
the electron-hole symmetry.
This is due to Kondo resonance and 
independent of the values of $v$ and $U$.
We also examine the effects of the asymmetry in the coupling 
$v_L^{\phantom{\dagger}} \neq v_R^{\phantom{\dagger}}$,
which disturbs the perfect transmission.
We note that preliminary results were 
reported in a proceedings.\cite{ao8}

In Sec.\ \ref{sec:MODEL}, 
we introduce the effective Hamiltonian,
and give the interpretation of the dc conductance 
and total charge displacement 
in terms of the free quasi-particles. 
In Sec.\ \ref{sec:e-h_symmetry}, 
we show some properties of the quasi-particles  
in the electron-hole symmetric case 
without specifying details of the model.
In Sec.\ \ref{sec:2nd_order}, 
we apply the method to the Hubbard chain of finite size, 
and present the results obtained with the second-order perturbation theory. 
Summary is given in Sec.\ \ref{sec:SUMMARY}.


\section{Effective Hamiltonian and ground-state properties}
\label{sec:MODEL}

In this section, 
we introduce an effective Hamiltonian for the free quasi-particles 
based on the perturbation theory in the inter-electron interaction. 
In this context, the dc conductance and total charge displacement 
at $T=0$ are described by the scattering coefficients of the  
free quasi-particles.
In what follows,
we will discuss the formulation assuming the single-mode leads 
for simplicity.  The formulation can be generalized 
to the multi-mode leads.

We start with a system which consists of three regions; 
a finite central region ($C$) 
and the two non-interacting leads on the left ($L$) and the right ($R$).
The central region consists of $N$ resonant levels, 
and the interaction is switched on only for the electrons in this region.
We assume that 
the two leads consist of infinite degrees of freedom, respectively,
and are connected to the central region 
by the mixing matrix elements $v_L^{\phantom{\dagger}}$ 
and $v_R^{\phantom{\dagger}}$ as illustrated in Fig.\ \ref{fig:single}. 
The complete Hamiltonian is given by 
\begin{eqnarray}
{\cal H} \ 
    &=&  \ {\cal H}_L + {\cal H}_R  + {\cal H}_C^0 + {\cal H}_C^{int}
            + {\cal H}_{mix} 
\label{eq:H}
\;, \\
{\cal H}_L &=&  \sum_{ij\in L} \sum_{\sigma} 
        \left(\,-t_{ij}^L - \mu\, \delta_{ij} \,\right)
           \,  c^{\dagger}_{i \sigma}\, c^{\phantom{\dagger}}_{j \sigma}
\;, 
\label{eq:H_L}
\\ 
{\cal H}_R  &=&  \sum_{ij\in R} \sum_{\sigma} 
        \left(\,-t_{ij}^R - \mu\, \delta_{ij} \,\right)
\,  c^{\dagger}_{i \sigma}\, c^{\phantom{\dagger}}_{j \sigma}
      \;, 
\label{eq:H_R}
\\      
  {\cal H}_C^{0} &=&   \sum_{ij\in C} \sum_{\sigma} 
        \left(\,-t_{ij}^C - \mu\, \delta_{ij} \,\right)\,
   c^{\dagger}_{i \sigma}\, c^{\phantom{\dagger}}_{j \sigma}
  \;, \\
    {\cal H}_C^{int} &=&   
{1 \over 2} \sum_{\{j\} \in C}\sum_{\sigma \sigma'}
    U_{j_4 j_3; j_2 j_1}\,  
 c^{\dagger}_{j_4 \sigma}\, c^{\dagger}_{j_3 \sigma'}
 c^{\phantom{\dagger}}_{j_2 \sigma'}\, c^{\phantom{\dagger}}_{j_1 \sigma}
  \;, 
\label{eq:H_int}
\\
  {\cal H}_{mix} &=& 
-  \sum_{\sigma} v_L^{\phantom{\dagger}} \left(\,  
             c^{\dagger}_{1 \sigma}\, c^{\phantom{\dagger}}_{0 \sigma}
             + 
          c^{\dagger}_{0 \sigma}\, c^{\phantom{\dagger}}_{1 \sigma}
 \,\right)
-  \sum_{\sigma} v_R^{\phantom{\dagger}} \left(\,  
             c^{\dagger}_{N+1 \sigma}\, c^{\phantom{\dagger}}_{N \sigma}
           + c^{\dagger}_{N \sigma}\, c^{\phantom{\dagger}}_{N+1 \sigma}
              \,\right)
  \;. 
\label{eq:H_mix}
\end{eqnarray}
Here $\, c^{\dagger}_{j \sigma}\,$ 
 ($c^{\phantom{\dagger}}_{j \sigma}$) creates (destroys) 
an electron with spin $\sigma$ at site $j$, 
and $\mu$ is the chemical potential.
$t_{ij}^{L}$, $t_{ij}^{R}$, and $t_{ij}^{C}$ are 
the intra-region hopping matrix elements
in each regions $L$, $R$, and $C$, respectively. 
The coupling between the central region and two leads
are described by the mixing matrix element  ${\cal H}_{mix}$. 
We assign the labels $1$, $2$, $\ldots$, $N$ to 
the sites in the central region.
Specifically, we assign the labels $1$ and $N$ to the sites 
at the interface on the left and right, respectively. 
Correspondingly, the label $0$ and $N+1$ are assigned 
to the sites at the lead-side of the interface on the left and right,
respectively [see Fig.\ \ref{fig:single}].
The inter-electron interaction 
$U_{j_4 j_3; j_2 j_1}$ is switched on in the central region, 
and it has the time reversal symmetry:  
$U_{4 3; 2 1}$ is real and 
$U_{4 3; 2 1}=U_{3 4; 1 2}=U_{1 2; 3 4 }=U_{4 2; 3 1}=U_{1 3; 2 4}$. 
We take all the hopping matrix elements to be real,
and will be using units $\hbar=1$ unless otherwise noted.

In the limit $N=1$, 
the model reduces to a single Anderson impurity 
in which the perturbation expansion is valid for all values 
of $U$.\cite{ZlaticHorvatic}
Our basic idea is to apply 
the perturbation theory in ${\cal H}_C^{int}$ following 
Yamada and Yosida\cite{YY1-2}
regarding the central region as one big impurity. 
Therefore, we take the unperturbed part of the Hamiltonian 
${\cal H}^{(0)}$ to be connected by including 
the mixing term ${\cal H}_{mix}$ in it; 
\begin{equation}
{\cal H}^{(0)}       
    =   {\cal H}_L + {\cal H}_R  
  + {\cal H}_C^0 + {\cal H}_{mix} \:.
\label{eq:H_0}  
\end{equation} 
Our working hypothesis is that 
the ground state is changed continuously 
against the adiabatic switching-on of the interaction ${\cal H}_C^{int}$.
The perturbation expansion can be done using   
the single-particle Green's function
\begin{equation} 
G_{jj'}(i\varepsilon_l) 
 =  
-    \int_0^{\beta} \! d\tau \,
   \left \langle \, T_{\tau} \,  
   c^{\phantom{\dagger}}_{j \sigma} (\tau) \, c^{\dagger}_{j' \sigma} (0)  
                  \,  \right \rangle  \, e^{i\, \varepsilon_l \tau} \;,
\label{eq:G_Matsubara}
\end{equation} 
where
$\beta= 1/T$, $\varepsilon_l = (2l+1)\pi/\beta$, 
$c_{j \sigma}(\tau) = e^{\tau  {\cal H}} c_{j \sigma} e^{- \tau  {\cal H}}$,
and $\langle \cdots \rangle$ denotes the thermal average 
$\mbox{Tr} \left[ \, e^{-\beta  {\cal H} }\, {\cdots}
\,\right]/\mbox{Tr} \, e^{-\beta  {\cal H} }$.
The spin index has been omitted from the left-hand side 
of Eq.\ (\ref{eq:G_Matsubara}) 
assuming the expectation value to be 
independent of whether spin is up or down.
Since the interaction is switched on only 
for the electrons in the central region,
the Dyson equation is written as
\begin{equation} 
  G_{ij}(z)    =   G^{(0)}_{ij}(z) 
    + \sum_{ll' \in C}\,G^{(0)}_{il}(z)\,  \Sigma_{ll'}(z)
   \, G_{l'j}(z) \;.
  \label{eq:Dyson}
\end{equation} 
Here $G^{(0)}_{ij}(z)$ is the unperturbed Green's function 
corresponding to ${\cal H}^{(0)}$.
The summations with respect to $l$ and $l'$ run over
the sites in the central region, 
and $\Sigma_{ll'}(z)$ is the self-energy correction
due to ${\cal H}_C^{int}$. 
Note that 
$G_{ij}(z) =  G_{ji}(z)$ and $\Sigma_{ij}(z) = \Sigma_{ji}(z)$ 
because of the time reversal symmetry of ${\cal H}$. 
In what follows, we will treat $z$ as a complex variable,
i.e.,  $G_{ij}(z)$ is the analytic continuation of $G_{ij}(i\varepsilon_l)$.
When the perturbation expansion is valid,
the single-particle excitation at the Fermi energy $z= i0^+$ 
does not decay at $T=0$ owing to a property 
of the Fermi liquid\cite{LangerAmbegaokar}  
\begin{equation} 
\mbox{Im}\,\Sigma_{ij}^+ (0)=0 \;.
\label{eq:nodamp} 
\end{equation} 
Here the superscript ($+$) is the label for a retarded function: we will use 
a notation $\Sigma_{ij}^{\pm}(\omega) \equiv \Sigma_{ij}(\omega \pm i0^+)\,$.
\cite{LangerAmbegaokar} 
Due to this property, 
the Dyson equation Eq.\ (\ref{eq:Dyson}) 
for $\omega=0$ can be mapped on to a scattering problem 
of the free quasi-particles described by the effective Hamiltonian
\begin{eqnarray}
 \widetilde{\cal H}_{qp}  \ &=& \  {\cal H}_L + {\cal H}_R 
                         + \widetilde{\cal H}_C + {\cal H}_{mix}
\label{eq:H_eff}
\;,\\
\widetilde{\cal H}_C \ &=& \   
\sum_{ij\in C}\sum_{\sigma} 
        \left(\,-\widetilde{t}_{ij}^C - \mu\, \delta_{ij} \,\right)
\, 
c^{\dagger}_{i \sigma}\, c^{\phantom{\dagger}}_{j \sigma} 
\;, 
\label{eq:H_eff_C} \\
 -\widetilde{t}_{ij}^C  \ &=& \  -t_{ij}^C \ 
\ + \ \mbox{Re}\,\Sigma_{ij}^{+}(0) 
\label{eq:t_eff} 
\;. 
\end{eqnarray}
The value of the Green's function corresponding 
to $\widetilde{\cal H}_{qp}$ and the one corresponding ${\cal H}$  are 
the same at $T=0$, $\omega=0$.

We now consider the conductance with the Kubo formalism.
If the ground state is a Fermi liquid, 
the contributions of the vertex corrections 
for the dc conductance vanish at $T=0$.\cite{ao5,ao6}
Therefore,  the dc conductance at $T=0$ is written 
in terms of the Green's function at $\omega=0$ also 
for interacting electrons;\cite{NOTE}
\begin{equation}
    g_N^{\phantom{\dagger}} \  = \ {2 e^2 \over h} \ 
    4\, 
        \Gamma_R(0) \, G_{N 1}^{+}(0) \,
    \Gamma_L(0) \,  G_{1 N}^{-}(0)   
        \;. 
\label{eq:cond}
\end{equation}
Here $\Gamma_{L}(\omega) = -\, 
\mbox{Im} 
\left[ v_{L}^2 \, \mbox{\sl g}_{L}^{+}(\omega)  \right]$, 
and $\Gamma_{R}(\omega) = -\, \mbox{Im} 
\left[ v_{R}^2 \, \mbox{\sl g}_{R}^{+}(\omega)  \right]$.
$\mbox{\sl g}_L^+$ ($\mbox{\sl g}_R^+$) is 
the Green's function at the interface $i=0$ ($i=N+1$) of the isolated lead,
and it is determined by ${\cal H}_L$ (${\cal H}_R$).
Note that Eq.\ (\ref{eq:cond}) is also expressed in terms of 
the transmission probability of the free quasi-particles, 
$|\widetilde{t}(0)|^2 = 
4 \Gamma_R(0) G_{N 1}^{+}(0) \Gamma_L(0) G_{1 N}^{-}(0) $.
\cite{Caroli,Landauer,Buttiker}
Correspondingly, the reflection probability 
is given by $\left|\widetilde{r}(0)\right|^2 =   
\left|1 -2i \Gamma_L(0)  G_{1 1}^{+}(0)\right|^2 
= \left|1 -2i \Gamma_R(0)  G_{N N}^{+}(0)\right|^2$,
and the unitarity $|\widetilde{t}(0)|^2 + |\widetilde{r}(0)|^2 = 1$ is 
preserved owing to the property Eq.\ (\ref{eq:nodamp}) 
[see also (\ref{eq:K})].
Another quantity which can be related to the scattering coefficients 
is the displacement of the total charge\cite{LangerAmbegaokar,Langreth} 
\begin{equation}
\Delta N_{tot}
 =   
\sum_{i\in C} \sum_{\sigma} 
 \langle c^{\dagger}_{i \sigma} c^{\phantom{\dagger}}_{i \sigma}
\rangle 
+ \sum_{i\in L} \sum_{\sigma} 
  \left[\, 
  \langle c^{\dagger}_{i \sigma} c^{\phantom{\dagger}}_{i \sigma} \rangle
  - 
  \langle c^{\dagger}_{i \sigma} 
  c^{\phantom{\dagger}}_{i \sigma} \rangle_{L}^{\phantom{0}}
    \,\right] 
+ \sum_{i\in R} \sum_{\sigma} 
  \left[\, 
  \langle c^{\dagger}_{i \sigma} c^{\phantom{\dagger}}_{i \sigma} \rangle
  - 
  \langle c^{\dagger}_{i \sigma} 
  c^{\phantom{\dagger}}_{i \sigma} \rangle_{R}^{\phantom{0}}
    \,\right] . 
\label{eq:dn_def}
\end{equation}
Here $\,\langle \cdots \rangle_{L}^{\phantom{0}}$  and 
$\,\langle \cdots \rangle_{R}^{\phantom{0}}$  
denote the ground-state average of isolated leads
described by ${\cal H}_{L}$ and ${\cal H}_{R}$, respectively.
At $T=0$, 
$\Delta N_{tot}$ can be expressed in terms of the $S$-matrix 
for the quasi-particles, following the derivation of 
the Friedel sum rule  by Langer and Ambegaokar,\cite{LangerAmbegaokar} 
as
\begin{eqnarray}
\Delta N_{tot}
&=& 
{1 \over \pi i}\  
\log [\, \det 
 \mbox{\boldmath $S$}
\,]  \;, 
\label{eq:Friedel} 
\\
\nonumber
\\
\mbox{\boldmath $S$}
 &=&  
  \left [ \,
 \matrix { 1  & 0   \cr
           0  & 1  \cr  }
 \, \right ]  
- \, 2\,i 
  \left [ \,
 \matrix { \Gamma_L(0)  & 0            \cr
           0            & \Gamma_R(0)  \cr  }
 \, \right ]  
  \left [ \,
 \matrix { G_{1 1}^{+}(0)  & G_{1 N}^{+}(0)  \cr 
           G_{N 1}^{+}(0)  & G_{N N}^{+}(0)  \cr  }
 \, \right ]  
     \;.   
\label{eq:S}
\end{eqnarray}
Therefore, 
the conductance and charge displacement are determined by
the inter- and intra-boundary elements 
of the Green's function such as $G_{N1}^{+}(0)$ and $G_{11}^{+}(0)$.

Next we discuss the structure of 
the Dyson equation Eq.\ (\ref{eq:Dyson}) further 
in order to make the mathematical features of the perturbation 
theory in the presence of the reservoirs clear. 
For the Green's functions in the central region, 
Eq.\ (\ref{eq:Dyson}) is written in a $N \times N$ matrix form; 
\begin{eqnarray}
\left\{\mbox{\boldmath ${\cal G}$}(z)\right\}^{-1}  
  &=&  
\{ \mbox{\boldmath ${\cal G}$}^{(0)}(z) \}^{-1}
- \mbox{\boldmath $\Sigma$}(z)   \;.
\label{eq:G_matrix}
\end{eqnarray}
Here $\mbox{\boldmath ${\cal G}$}(z) = \{G_{ij}(z)\}$ with $ij \in  C$,
and the inverse matrix of the unperturbed part can be 
expressed as $\{ \mbox{\boldmath ${\cal G}$}^{(0)}(z) \}^{-1}=
z \, \mbox{\boldmath $1$} 
-  \mbox{\boldmath ${\cal H}$}_C^0  
- \mbox{\boldmath ${\cal V}$}_{mix}(z)$   with 
\begin{eqnarray}
 \mbox{\boldmath ${\cal H}$}_C^0 \ &=& \  
\left [ \,
 \matrix { -t_{11}^C-\mu    & -t_{12}^C      & \cdots  &               \cr
            -t_{21}^C        & -t_{22}^C -\mu &         &               \cr
            \vdots          &               & \ddots  &               \cr 
                            &               &         & -t_{NN}^C -\mu \cr
          }
          \, \right ]  
\;,
\\
\nonumber
\\
\mbox{\boldmath ${\cal V}$}_{mix}(z) \  &=& \ 
 \left [ \, \matrix { 
 v_L^{2} \, \mbox{\sl g}_L(z) & 0 & \cdots &0 & 0 \cr
 0           &    0     & \cdots   &  0 & 0     \cr
 \vdots      &  \vdots  & \ddots   &  \vdots & \vdots  \cr 
 0 & 0 & \cdots & 0 & 0 \cr
 0 & 0 & \cdots & 0 & v_R^{2} \, \mbox{\sl g}_R(z) \cr
                    }
 \, \right ]  
 \;, 
\label{eq:V_mix} \\
 \mbox{\boldmath $\Sigma$}(z) \  &=&  \  
\left [ \,
 \matrix {  \Sigma_{11}(z)    & \Sigma_{12}(z)  & \cdots  &             \cr
            \Sigma_{21}(z)    & \Sigma_{22}(z)  &         &             \cr
            \vdots      &                 & \ddots  &             \cr 
                        &                 &         & \Sigma_{NN}(z) \cr
          }
          \, \right ]  
\;.
\end{eqnarray}
Here $\mbox{\boldmath $1$}$ is the $N \times N$ unit matrix.
$\mbox{\boldmath ${\cal V}$}_{mix}$ corresponds to 
the contribution of the mixing with the reservoirs. 
Especially, the two non-zero elements 
 $v_{L}^2 \, \mbox{\sl g}_{L}$ and $v_{R}^2 \, \mbox{\sl g}_{R}$
have the finite imaginary parts corresponding 
to $\Gamma_L$ and $\Gamma_R$ defined just bellow Eq.(\ref{eq:cond}).
These imaginary parts change the discrete levels 
of $\mbox{\boldmath ${\cal H}$}_C^0$ 
to the continuous peaks with finite level width, and thus 
the unperturbed part of the Green's 
function $\mbox{\boldmath ${\cal G}$}^{(0)}(z)$  describes 
a system of $N$ resonant scatterers.
Therefore, owing to the contribution of the mixing, 
the mathematical structure of the perturbation theory 
in the presence of the reservoirs 
becomes similar to that of the Anderson model with a number of orbits 
rather than the usual Hubbard model without reservoirs.
In this matrix form, the quasi-particle description is summarized as follows. 
Due to the 
property $\mbox{Im}\, \mbox{\boldmath $\Sigma$}^{+} (0) =0$ at $T=0$, 
the Green's function at $\omega=0$ is written as   
$\left\{\mbox{\boldmath ${\cal G}$}^+(0)\right\}^{-1}  = 
\mbox{\boldmath ${\cal K}$}  
-  \mbox{\boldmath ${\cal V}$}_{mix}^+(0)$, where  
$\mbox{\boldmath ${\cal K}$}$  is 
the renormalized hopping matrix 
corresponding to $-\widetilde{\cal H}_C$ defined 
by Eq.\ (\ref{eq:H_eff_C});
\begin{eqnarray}
\mbox{\boldmath ${\cal K}$} &=&   
- \left[\, 
   \mbox{\boldmath ${\cal H}$}_C^0  
            + \mbox{Re}\, \mbox{\boldmath $\Sigma$}^+(0)
  \,\right]
  . 
\label{eq:K} 
\end{eqnarray}
Then a relation corresponding to the optical theorem can be obtained as  
$
\mbox{\boldmath ${\cal G}$}^+(0) - 
\mbox{\boldmath ${\cal G}$}^-(0)
 = 
\mbox{\boldmath ${\cal G}$}^+(0) 
 \left[ 
  \mbox{\boldmath ${\cal V}$}_{mix}^+(0)
   -   
  \mbox{\boldmath ${\cal V}$}_{mix}^-(0)  
 \right] 
 \mbox{\boldmath ${\cal G}$}^-(0)
$, which certifies the unitarity of the scattering coefficients. 
Here the superscript $+$ ($-$) means 
the retarded (advanced) function.

The quasi-particle approach cannot be applied 
to a non-Fermi-liquid ground state where   
the perturbation theory with respect 
to the inter-electron interaction breaks down.
However, 
in the finite interacting system connected to reservoirs 
as illustrated in Fig.\ \ref{fig:single},
a cross over from a high temperature phase 
to the low temperature Fermi liquid phase is expected in many cases.
This is because, due to the connection with the reservoirs, not only 
the energy scale corresponding to the level spacing of the isolated 
interacting system of size $N$ but also the level width 
introduced through $\Gamma_L$ and $\Gamma_R$, i.e., 
the imaginary part of the mixing term Eq.\ (\ref{eq:V_mix}),  
plays a role of the cut-off. 
In this sense, as mentioned above, the system can be regarded as  
a generalized Anderson impurity with $N$ resonant states.

When the interacting region is described by an one-dimensional chain, 
the Tomonaga-Luttinger (TL) behavior\cite{Solyom} seems 
to be seen for large $N$. 
However, as discussed by Kane and Fisher
for the finite TL model of size $L$ connected 
to Fermi liquid reservoirs,\cite{KaneFisher} 
the cross over from the high temperature TL phase to
the low temperature Fermi liquid phase occurs at 
a characteristic energy $T_0 \simeq v_F/L$,
where $v_F$ is the Fermi velocity.
Since this simple discussion does not take into account 
various effects such as the back and Umklapp scatterings, 
the characteristic energy will be different 
depending on the details of the situations.
Specifically, in the electron-hole symmetric case 
of the Hubbard chain examined in the following chapters, 
the characteristic energy depends on whether $N$ is even or odd, 
and it should be the Mott-Hubbard gap $E_G$ or 
the Kondo temperature $T_K$.
Again, due to the contribution of the mixing,
the low energy excitations 
below the characteristic energy scale 
are described by the Fermi liquid 
as far as $N$ is finite.
For even $N$, the Mott-Hubbard gap opens 
in the thermodynamic limit of the usual Hubbard model without reservoirs,
and the corresponding tendency should be seen in the present system 
of finite $N$. 
However, there still remains a finite density of states 
at the Fermi energy caused by the mixing with the reservoirs,
although the spectral weight in the 
region corresponding to the gap will decrease with increasing $N$.
For odd $N$, there is an additional spectral weight 
at the Fermi energy in the Mott-Hubbard gap, i.e.,
the Kondo resonance of the width $T_K$. 
Thus, the insulator-like low-conductivity should be seen 
at the temperature range $T_K<T<E_G$, 
and the metallic behavior of the Fermi liquid will be seen 
at low temperatures $T<T_K$. 
Note that $T_K$ can be defined only for the open system 
connected to reservoirs, while $E_G$ is a constant defined 
in the thermodynamic limit of the isolated system. 
Furthermore, $T_K$ should decrease with increasing $N$,
and in most of cases $T_K$ should vanish in the limit of $N \to \infty$. 
Therefore, if the limit $N \to \infty$ is taken keeping $T$ to be finite,
the even-odd behavior disappears and 
at $T<E_G$ the insulating behavior 
meets with an intuitive picture will be seen.
This limit corresponds to the thermodynamic limit 
of the macroscopic system.  
However, there is another nontrivial limit which describes 
the low temperature physics of the mesoscopic systems, i.e.,
the limit $T \to 0$  keeping $N$ to be finite.
The ground state of the small system belongs to this limit, 
and in the case of the small Hubbard chain 
the even-odd behavior can be seen at low temperatures 
since $T_K$ is finite for small $N$. 
This kind of a mesoscopic limit is expected to be realized 
in quantum dots or quantum wire of nanometer size,
and our aim here is to develop a Fermi-liquid theory  
for the mesoscopic systems.

Another possible non-Fermi-liquid state is the ground state 
of the multi-channel Kondo system.\cite{CoxZawadowski} 
Shimizu, Sakai and Suzuki have shown 
with the numerical renormalization group approach 
that the non-Fermi-liquid ground state is realized 
also in an extended version of the Anderson model.\cite{ShimizuY}
So, if the parameters of the Hamiltonian Eq.\ (\ref{eq:H}) satisfy 
certain conditions, the similar situation seems to be realized.


\section{Quasi-particles in the electron-hole symmetric case}
\label{sec:e-h_symmetry}

The effective Hamiltonian 
has some notable properties in the electron-hole symmetric case 
where the average number of electrons in each site is unity. 
In this section, 
we provide a simplified expression of the  conductance 
in the electron-hole symmetric case. 
The results Eqs.\ (\ref{eq:g_1_eff}) and (\ref{eq:g_2_eff}) 
will be used for the finite Hubbard chain in the next section. 
Especially, form Eq.\ (\ref{eq:g_1_eff}), 
we can deduce quite generally that
the perfect transmission occurs for odd $N$ 
when the system has an additional inversion symmetry.

In the electron-hole symmetric case,
the off-diagonal element 
of the renormalized parameter $\widetilde{t}_{ij}^C$ is zero 
when  $i$ and $j$ belong to the same sublattice, i.e., 
$-\widetilde{t}_{ij}^C - \mu\, \delta_{ij}=0\,$  
for $|i-j| =0,\, 2,\, 4,\, \ldots$,
and $\mbox{Re}\, \mbox{\boldmath ${\cal V}$}_{mix}^+(0) =0$.    
Thus, the matrix $\mbox{\boldmath ${\cal K}$}$  
defined by Eq.\ (\ref{eq:K}) 
has a checkered structure, and    
the Green's function at $T=0$, $\omega=0$  is written in the form;  
\begin{eqnarray}
\left\{\mbox{\boldmath ${\cal G}$}^+(0)\right\}^{-1}  = \ 
\left [ \, 
\matrix {  
 0  & \widetilde{t}_{12}^C  & 0 & \widetilde{t}_{14}^C &  \cdots \cr
 \widetilde{t}_{21}^C & 0  & \widetilde{t}_{23}^C      & 0 &     \cdots \cr
 0  & \widetilde{t}_{32}^C  & 0 & \widetilde{t}_{34}^C & \cdots \cr
 \widetilde{t}_{41}^C & 0  & \widetilde{t}_{43}^C      & 0 &  \cdots \cr
 \vdots             & \vdots  & \vdots & \vdots & \ddots \cr
        }
\, \right ]  
\ + \ \left [ \,
 \matrix {  i \Gamma_L(0) &           &               \cr
                       &  \mbox{\boldmath $0$} &   \cr
                       &           &  i \Gamma_R(0)   \cr
         }
          \, \right ] \;,  
\label{eq:G_eff_sy} 
\end{eqnarray}
and $G_{ij}^{+}(0)$ is obtained by taking 
the inverse of Eq.\ (\ref{eq:G_eff_sy}). 
Specifically, using the explicit form of $G_{1N}^{+}(0)$, 
the dc conductance $g_N^{\phantom{\dagger}}$ can 
be obtained from Eq.\ (\ref{eq:cond}). 
For $N=1$ and $2$, the dc conductance is written in the form 
\begin{eqnarray}
 g_1^{\phantom{\dagger}} &=&  {2 e^2 \over h} \ 
           { \Gamma_L \,\Gamma_R   
             \over
            \left[\, (\Gamma_L + \Gamma_R)/2 \,\right]^2} \;, 
\label{eq:g_1}
\\
\nonumber \\
 g_2^{\phantom{\dagger}} &=&  {2 e^2 \over h} \ 
     { \Gamma_L \,\Gamma_R\  \{\widetilde{t}_{12}^C \}^{2}_{\phantom{0}} 
             \over
\left[\, \left( \Gamma_L \, \Gamma_R  
+ \{\widetilde{t}_{12}^C\}^{2}_{\phantom{0}}\right)/2 \,\right]^2} \;.
\label{eq:g_2}
\end{eqnarray}
Here $\Gamma_{L} \equiv \Gamma_{L}(0)$, $\Gamma_{R} \equiv \Gamma_{R}(0)$,
and we will use this simplified notation in what follows.  
For $N \geq 3$,  $G_{1N}^{+}(0)$ is written in the form
\begin{eqnarray}
& &  
G_{1N}^{+}(0) \ = \ 
                 (-1)^{N+1} \, {  \det \mbox{\boldmath ${\cal K}$}_{N1}
                     \over
                \det \left\{\mbox{\boldmath ${\cal G}$}^+(0)\right\}^{-1}  
                   }
\label{eq:G1N} 
\;, 
\\
\nonumber \\
& & 
\det \left\{\mbox{\boldmath ${\cal G}$}^+(0)\right\}^{-1}  
\, = \ 
\left\{ 
       \begin{array}{ll}
         i\, \Gamma_L \, \det \mbox{\boldmath ${\cal K}$}_{11} + 
         i\, \Gamma_R \, \det \mbox{\boldmath ${\cal K}$}_{NN} \;,  & 
        \  \mbox{for \ odd \ $N$}  \\
     - \Gamma_L \,\Gamma_R \, \det \mbox{\boldmath ${\cal K}$}_{11}^{NN} 
     + \det \mbox{\boldmath ${\cal K}$} \;,  & 
        \  \mbox{for \ even \ $N$}  
        \end{array} \right.   \;.
\label{eq:det_Ginv}
\end{eqnarray}
Here $\mbox{\boldmath ${\cal K}$}_{ij}$ is 
a $(N-1) \times (N-1)$ matrix obtained from $\mbox{\boldmath ${\cal K}$}$ 
by deleting the $i$-th row and the $j$-th column.
Similarly, $\mbox{\boldmath ${\cal K}$}_{11}^{NN}$ is 
a $(N-2) \times (N-2)$ matrix obtained from $\mbox{\boldmath ${\cal K}$}$ 
by deleting the first and the $N$-th rows, 
and the first and the $N$-th columns.
Due to the checkered structure of Eq.\ (\ref{eq:G_eff_sy}), 
the determinants satisfy a relation  
\begin{equation}
(\det \mbox{\boldmath ${\cal K}$}_{N1})^2 =  
\left\{ \begin{array}{cl}
\det \mbox{\boldmath ${\cal K}$}_{11}
\, \det \mbox{\boldmath ${\cal K}$}_{NN} \;, & 
        \ \mbox{for \ odd \ $N$}  \\
-\det \mbox{\boldmath ${\cal K}$}
\, \det \mbox{\boldmath ${\cal K}$}_{11}^{NN} \;, &
         \ \mbox{for \ even \ $N$} 
    \end{array} \right. \;.
\label{eq:K_relation}
\end{equation}
Although the details are given in the appendix \ref{sec:symmetric_app}, 
the origin of the even-odd dependence 
can be understood just seeing typical examples; 
\begin{eqnarray}
\mbox{\boldmath ${\cal K}$}  \ &=& \ 
\left [ \, 
\matrix {  
 0  & \widetilde{t}_{12}^C  & 0 & \widetilde{t}_{41}^C & 0 \cr
 \widetilde{t}_{12}^C & 0 & \widetilde{t}_{23}^C & 0 & \widetilde{t}_{52}^C \cr
 0  & \widetilde{t}_{23}^C & 0 & \widetilde{t}_{34}^C & 0  \cr
 \widetilde{t}_{41}^C & 0 & \widetilde{t}_{34}^C & 0 & \widetilde{t}_{45}^C \cr
 0  & \widetilde{t}_{52}^C  & 0 & \widetilde{t}_{45}^C & 0  \cr
        }
\, \right ]  
  \qquad  \mbox{for $\,N=5$}  \;,
\label{eq:K5}
\\
\nonumber \\
\nonumber \\
\mbox{\boldmath ${\cal K}$} \ &=& \ 
\left [ \, 
\matrix {  
0 & \widetilde{t}_{12}^C &0& \widetilde{t}_{41}^C &0& \widetilde{t}_{61}^C \cr
\widetilde{t}_{12^C} &0& \widetilde{t}_{23}^C  &0& \widetilde{t}_{52}^C &0  \cr
0 & \widetilde{t}_{23}^C &0& \widetilde{t}_{34}^C &0& \widetilde{t}_{63}^C \cr
\widetilde{t}_{41}^C &0 & \widetilde{t}_{34}^C &0& \widetilde{t}_{45}^C &0  \cr
0 & \widetilde{t}_{52}^C &0& \widetilde{t}_{45}^C &0& \widetilde{t}_{56}^C \cr
\widetilde{t}_{61}^C &0& \widetilde{t}_{63}^C &0& \widetilde{t}_{56}^C &0  \cr
        }
\, \right ]      \qquad  \mbox{for $\,N=6$}  
\;.
\label{eq:K6}
\end{eqnarray}
The antidiagonal elements, 
${\cal K}_{j,\,N+1-j}\,$  for $j=1,\,2,\,\ldots,\, N$, 
are zero for odd $N$, 
while those are not for even $N$.  
Furthermore, 
$\det \mbox{\boldmath ${\cal K}$}$ is zero for odd $N$, 
while it is finite and can be divided into two determinants 
for even $N$. 
Using the properties Eqs.\ (\ref{eq:G1N})--(\ref{eq:K_relation}) 
with Eq.\ (\ref{eq:cond}),
the dc conductance for odd $N$ ($=2M+1$) is written 
in the form   
\begin{eqnarray}
 g_{2M+1}^{\phantom{\dagger}} &=&  {2 e^2 \over h} \ 
           { \widetilde{\Gamma}_L \,\widetilde{\Gamma}_R   
             \over
            [\, (\widetilde{\Gamma}_L + \widetilde{\Gamma}_R)/2 \,]^2} \;, 
\label{eq:g_1_eff}
\end{eqnarray}
where $\widetilde{\Gamma}_L = \lambda\,\Gamma_L$,    
$\widetilde{\Gamma}_R  = \Gamma_R \,/\, \lambda$, and 
\begin{eqnarray}
 \lambda &=&   \sqrt{ 
                      \det \mbox{\boldmath ${\cal K}$}_{11} \over
                      \det \mbox{\boldmath ${\cal K}$}_{NN} 
                    } \;.
\label{eq:lambda}
\end{eqnarray}
We note that the parameter $\lambda$ 
can be simplified as Eq.\ (\ref{eq:lambda_2}) 
using the checkered structure further.
If the system also has 
the inversion symmetry $\Gamma_L=\Gamma_R$ 
in addition to the electron-hole symmetry, 
the parameter is fixed as $\lambda =1$. 
This is because the matrix $\mbox{\boldmath ${\cal K}$}$ becomes 
symmetric with respect to the antidiagonal line.
Therefore, due to the combination of the symmetries, 
the perfect transmission occurs $g_{2M+1}^{\phantom{\dagger}}={2 e^2 / h}$  
independent of $M$ and the details of the interaction, 
as far as the perturbation expansion is valid.
Physically, this is caused by the Kondo resonance  
appearing at the Fermi energy $\omega=0$.
On the other hand, for even $N$ ($=2M$), 
the dc conductance is written  in the form 
\begin{eqnarray}
 g_{2M}^{\phantom{\dagger}} &=&  {2 e^2 \over h} \ 
           { \Gamma_L \,\Gamma_R\  \widetilde{v}_{C}^{\,2} 
             \over
\left[\, \left(\Gamma_L \Gamma_R + \widetilde{v}_{C}^{\,2}
              \right)/2 \,\right]^2} \;, 
\label{eq:g_2_eff}
\\
\nonumber 
\\
\widetilde{v}_{C}^{\,2} &=&  
- {\det \mbox{\boldmath ${\cal K}$} \over
   \det \mbox{\boldmath ${\cal K}$}_{11}^{NN}} 
\;.
\label{eq:v_eff}
\end{eqnarray}
We note that the parameter $\widetilde{v}_{C}$
can also be simplified as Eq.\ (\ref{eq:v_eff_2}). 
These two expressions for the dc conductance, 
Eqs.\ (\ref{eq:g_1_eff}) and (\ref{eq:g_2_eff}),
are the main results of this section.

We next examine the Friedel sum rule.
Using the property Eq.\ (\ref{eq:det_Ginv}) and 
the expression of the total charge displacement 
Eqs.\ (\ref{eq:Friedel}) and (\ref{eq:S}), 
we obtain $\det \mbox{\boldmath $S$}=1$ for even $N$, 
and $\det \mbox{\boldmath $S$}=-1$ for odd $N$. 
This is consistent with the fact that 
the average number of electrons in each site is unity. 
In order to realize the electron-hole symmetry, 
the bare matrix element $t_{ij}^{C}$ should also 
have the checkered structure as that is assumed for
the renormalized parameter $\widetilde{t}_{ij}^C$,
i.e., the system must be classified into two sublattices.
Note that, since only the symmetry is assumed so far,
the bare hopping matrix element is not necessary to 
be restricted to the nearest-neighbor one.
Furthermore, the system is possible to be disordered 
through the randomness in the off-diagonal element of $t_{ij}^C$.


\section{Transport through a finite Hubbard chain}
\label{sec:2nd_order}

In this section, we will use the effective Hamiltonian 
to investigate the conductance of a finite Hubbard chain 
connected to two semi-infinite leads. 
The system is considered as a model for a series 
of quantum dots or atomic wire of nanometer size.
We calculate the renormalized 
hopping matrix element $\widetilde{t}_{ij}^{C}$ up to the second order in $U$,
and obtain the dc conductance using the expressions given 
in the previous section.
We note that the second-order perturbation theory has been used 
by several groups for studying transport properties  
of a single impurity\cite{HDW2,YMF,MiiMakoshi,TakagiSaso}
and systems consisting of a number of resonant levels.\cite{PFA,KKNO}
For $N \geq 2$, the off-diagonal part of the self-energy 
plays an important role on the conductance at $T=0$ as seeing
in the results presented below. 
To our knowledge, however, effects of the off-diagonal part 
have not been examined sufficiently so far.

The schematic picture of the model is illustrated in Fig.\ \ref{fig:lattice}: 
the system consists of $N$ interacting sites
at the center $\,1 \leq i \leq N$, 
and two  non-interacting leads 
at $\,-\infty < i \leq 0\,$ and $\,N+1 \leq i < +\infty$.  
The explicit form of the Hamiltonian is given by 
\begin{eqnarray}
H &=&  H_0 + H_I \;,
\label{eq:H_Hubbard} \\
H_0 &=& 
 - \sum_{i=-\infty}^{+\infty} \sum_{\sigma}\,
 t_{i}^{\phantom{\dagger}}
  \left (\,  c^{\dagger}_{i+1 \sigma}\, c^{\phantom{\dagger}}_{i \sigma}
     \, + \, c^{\dagger}_{i \sigma}\, c^{\phantom{\dagger}}_{i+1 \sigma}
     \, \right ) 
 - \mu \sum_{i=-\infty}^{+\infty} \sum_{\sigma} 
    c^{\dagger}_{i \sigma}\, c^{\phantom{\dagger}}_{i \sigma}
\nonumber \\     
     & &
   + \  \sum_{i=1}^{N} \sum_{\sigma}  
     \left(\epsilon_0 + {U \over 2} \right) n_{i \sigma}^{\phantom{\dagger}}
\;, 
\label{eq:H_0_hub} 
\\
H_I &=& U  \sum_{i=1}^{N} \left[\, 
                 n_{i \uparrow}^{\phantom{\dagger}}
             \,n_{i \downarrow}^{\phantom{\dagger}}
- {1 \over 2}  \,  (   n_{i \uparrow}^{\phantom{\dagger}}
                   + n_{i \downarrow}^{\phantom{\dagger}} \, )
            \,\right]
\;,
\label{eq:H_I_hub} 
\end{eqnarray}
where $U$ is the repulsive interaction in the chain, $\epsilon_0$ is 
the on-site energy, 
and $n_{i \sigma}^{\phantom{\dagger}} =
 c^{\dagger}_{i \sigma}\, c^{\phantom{\dagger}}_{i \sigma}$. 
We take the hopping matrix element 
to be $t_i^{\phantom{\dagger}} = t$ for every links except for 
the coupling between the chain and leads, 
i.e.,  $t_0^{\phantom{\dagger}}=v_L^{\phantom{\dagger}}$ 
and $t_N^{\phantom{\dagger}}=v_R^{\phantom{\dagger}}$. 
Thus, the level width caused by the coupling with the leads  
is given by $\Gamma_{\alpha}(0) =  \pi \, v_{\alpha}^2 D(0)$ 
for $\alpha=L,\,R$ with $D(0)=\sqrt{4t^2-\mu^2}\,/\,(2 \pi t^2)$. 
In what follows, we concentrate on the electron-hole symmetric case  
taking the parameters to be $\mu=0$ and $\epsilon_0=-U/2$.

Within the second order, 
the self-energy correction is described by a diagram Fig.\ \ref{fig:diagram},
and the value at $T=0$, $\omega=0$ is obtained from the expression
\begin{eqnarray}
 \Sigma^{(2)}_{ij}(i0^+) 
    =  - U^2   \int_{-\infty}^{\infty} \! \int_{-\infty}^{\infty} 
         \frac{d\epsilon\, d\epsilon'}{(2\pi)^2}
       \, G^{(0)}_{ij}(i \epsilon)  \,  
          G^{(0)}_{ij}(i \epsilon') \, 
          G^{(0)}_{ji}(i \epsilon + i \epsilon') 
\;,
 \label{eq:Self_2v} 
\end{eqnarray}
for $1\leq i,\, j \leq N$.
Here $G^{(0)}_{ij}(i \epsilon)$ is the unperturbed Green's 
function corresponding to $H_0$, 
and thus  $\Sigma^{(2)}_{ij}(i0^+)$ depends on 
$v_L^{\phantom{\dagger}}$ and $v_R^{\phantom{\dagger}}$ 
through $G^{(0)}_{ij}$. 
Note that the value of the retarded function at $\omega=0$ 
can be obtained from the Matsubara function, i.e.,   
$\Sigma_{ij}^{(2)}(i0^+) = \Sigma_{ij}^{(2)}(i\epsilon)|_{\epsilon \to 0^+}$.
The imaginary part of the self-energy vanishes 
owing to the Fermi-liquid property, 
i.e., $\mbox{Im} \Sigma^{(2)}_{ij}(i0^+)=0$. 
Furthermore, $\Sigma^{(2)}_{ij} (i0^+) = 0$ for even $|i-j|$  
since $G_{ij}^{(0)}(-i\epsilon) = (-1)^{|i-j|+1} \,G_{ij}^{(0)}(i\epsilon)$ 
in the electron-hole symmetric case.   
We have done the integration Eq.\ (\ref{eq:Self_2v}) numerically 
to estimate the renormalized matrix 
element $\widetilde{t}_{ij}^{C}$ defined by 
Eq.\ (\ref{eq:t_eff}) within the second order, and obtain 
the dc conductance $\,g_N^{\phantom{\dagger}}$ 
using Eqs.\ (\ref{eq:g_1_eff}) and (\ref{eq:g_2_eff}). 
As an example, we show the result of the self-energy for $N=6$, 
\begin{eqnarray}
    \mbox{\boldmath $\Sigma$}^{(2)}(i0^+)  &=& \, -t\, 
 \left({U \over 2\pi t} \right)^2 \, 
\widetilde{\mbox{\boldmath $\Sigma$}}^{(2)} 
\;, \\
\nonumber \\
\widetilde{\mbox{\boldmath $\Sigma$}}^{(2)} 
\  &=&
\left[\, 
\matrix{
  0&0.8596&0&-0.1642&0&0.0783 \cr
  0.8596&0&0.3969&0&-0.0352&0 \cr
  0&0.3969&0&0.7543&0&-0.1642 \cr
  -0.1642&0&0.7543&0&0.3969&0 \cr
  0&-0.0352&0&0.3969&0&0.8596 \cr
  0.0783&0&-0.1642&0&0.8596&0 \cr
}
\,\right] \;.
\label{eq:self_N=6}
\end{eqnarray}
Here the coupling is taken to be 
$v_L^{\phantom{\dagger}} /t = v_R^{\phantom{\dagger}} /t =0.8$.
The matrix element for a given distance $|i-j|$ shows an oscillatory behavior,
i.e., $\Sigma_{l,l+2m+1}^{(2)}(i0^+)$ is an oscillatory function of $\,l$.
In contrast, 
in the special case $v_L^{\phantom{\dagger}} = v_R^{\phantom{\dagger}} = t$, 
$\Sigma_{l,l+2m+1}^{(2)}(i0^+)$ is independent of $l$ and 
the integration Eq.\ (\ref{eq:Self_2v}) can be done analytically.\cite{ao7}
Note that $\Sigma_{ij}^{(2)}(i0^+)$ is also 
an oscillatory function of $|i-j|$,  
and the absolute value tends to decrease with increasing $|i-j|$.

\subsection{
Symmetric connection $\,v_L = v_R$ 
}
\label{subsec:sym_LR}

We now examine the case
where the system has the inversion symmetry,  
$v_L^{\phantom{\dagger}} = v_R^{\phantom{\dagger}} \equiv v$,  
in addition to the electron-hole symmetry. 
In this case, as it was shown generally in the previous section,  
the perfect transmission occurs for odd $N$ independent of $U$. 
Physically, this is due to 
the contribution of the Kondo resonance appearing at the Fermi energy.
So, we have done the numerical calculation only for even $N$.
The result of $g_N^{\phantom{\dagger}}$ is 
plotted as a function of the size $N$ in Fig.\ \ref{fig:g_N_symm}, 
where the parameters are taken to be $v/t =0.8$ 
and $U/(2\pi t) =0.4$. 
The result shows a typical even-odd oscillatory behavior,
and the conductance for even $N$ decreases with increasing $N$.  
This reduction is caused by the oscillatory $l$ dependence of 
the self-energy $\Sigma_{l,l+2m+1}^{(2)}$.
In Fig.\ \ref{fig:g_2M_symm_several_U},    
the envelope of $g_N^{\phantom{\dagger}}$ for even $N$ ($=2M$) is 
plotted vs $M$ for several values of 
the repulsion $U/(2\pi t) =0.0,\,0.2,\,0.4,\,0.6$, and $0.8$,  
taking the mixing matrix element to be $v/t  = 0.8$.  
The conductance $g_{2M}^{\phantom{\dagger}}$ 
decreases with increasing $M$ and increasing $U$. 
In the non-interacting case, $g_{2M}^{\phantom{\dagger}}$ is 
independent of $M$ since  
the parameter defined by Eq.\ (\ref{eq:v_eff}) is 
simply $\widetilde{v}_C = t$.  
The reduction of the conductance 
seems to show  a tendency toward a Mott-Hubbard insulator.
However, the result shows a rather moderate decay, i.e., 
it does not show an exponential dependence. 
Since the range of $U$ in which 
the second-order perturbation theory is able to provide 
quantitatively reliable results depends on the size $M$,
contributions of higher-order terms 
should be examined in order to clarify 
the correct asymptotic behavior for large $M$.

In Fig.\ \ref{fig:g_2M_symm_U}, 
the conductance is plotted as a function of $U$ for 
a number of even $N$ ($=2,\,4,\,6,\,\ldots$) 
taking the parameters to be $v/t=0.8$ (dashed lines) 
and $v/t=1.0$ (solid lines).
As it can be seen in the behavior of the dashed lines, 
the reduction of $g_{2M}^{\phantom{\dagger}}$ is 
proportional to $U^2$ for small $U$ when $|v/t| < 1.0$.  
The curvature increases with the size $M$. 
In order to see this quantitatively, 
we expand the conductance in powers 
of $U$ as $g_{2M}^{\phantom{\dagger}}= (2\,e^2/h)
[ C_0 - C_2 \{U/(2\pi t)\}^2 + \cdots]$,
and plot the ratio $C_2/C_0$ as a function of $M$ 
in Fig.\ \ref{fig:c2_c0} for 
several values of $v/t$ ($=0.1,\, 0.2,\, \ldots,\, 0.9 $). 
$C_2$ increases with the size $M$, 
and seems to diverge in the limit of large $M$ showing 
a power low behavior. 
Note that the transmission probability 
in the non-interacting case $C_0$ depends on 
the mixing matrix element $v$ but is independent 
of the size $M$ as mentioned in the above.
Furthermore, since the self-energy is calculated up to the order $U^2$,
the result of $C_2$ is exact. 
The ratio $C_2/C_0$ increases with decreasing $v$.
This means that the effect of the interaction 
is enhanced when the coupling between the sample and leads is weak.
In the special case $v = t$, 
the reduction of $\,g_{2M}^{\phantom{\dagger}}$ is proportional 
to $U^4$ for small $U$, 
i.e., $g_{2M}^{\phantom{\dagger}}= (2\,e^2/h)[ 1 - C_4 \{U/(2\pi t)\}^4 
+ \cdots ]$.
This is because the unperturbed system 
described by $H_0$ has a translational invariance 
accidentally in this case 
and the reflection probability is zero at $U=0$.
The solid lines for $N \gtrsim 10$ in Fig.\ \ref{fig:g_2M_symm_U} 
are almost overlapping each other.
In Fig.\ \ref{fig:c4}, $C_4$ is plotted vs $M$.
This result is also exact, and the coefficient $C_4$ converges 
to a finite value for large $M$; $C_4 \simeq 0.5293$ for $N=250$. 
In the appendix \ref{sec:reflection},
the convergence of $C_4$ in the limit of $M \to \infty$ is 
confirmed with another approach. 
Therefore, when $v=t$, $g_{2M}^{\phantom{\dagger}}$ is 
finite even in the limit of large $M$ at least for small $U$.\cite{ao7} 
For large values of $U$, both the dashed and solid lines 
in Fig.\ \ref{fig:g_2M_symm_U} tends to zero showing a $1/U^{4}$ dependence. 
However this behavior depends on the approximation: 
as it can be confirmed from Eqs.\ (\ref{eq:g_2_eff}) 
and (\ref{eq:v_eff}), $\widetilde{v}_C \propto U^n$ 
and $g_{2M}^{\phantom{\dagger}} \propto 1/U^{2n}$ 
for large $U$ when the self-energy is estimated 
within the $n$-th order perturbation in $U$.

The mixing matrix element $v$ determines the bare level width 
of the resonant states, 
and in the present case it is given by $\Gamma = v^2/t$. 
Since we are now considering the electron-hole symmetric case, 
the Fermi level for even $N$ is located between the two resonant states 
corresponding to the highest occupied and lowest unoccupied levels.
In Fig.\ \ref{fig:g_2M_symm_U2},  $g_{2M}^{\phantom{\dagger}}$ is 
plotted vs $U$ for several 
values of  $v/t$ ($=0.4,\,0.5,\,0.6,\,\ldots,\,1.0$), 
where the solid (dashed) lines are the results 
for $N=40$ ($N=4$).
The value of the conductance itself decreases with $v$,  
but the qualitative feature of the $U$ dependence 
is similar in the cases of $|v/t| <1.0$.
In Fig.\ \ref{fig:g_2M_symm_several_V}, 
we show the conductance as a function of $M$ 
for several $v/t$ ($=0.4,\,0.5,\,\ldots,\,1.0$) 
taking the repulsion to be $U/(2\pi t) =0.3$. 
In this figure, the conductance is normalized by 
the non-interacting value $g_{2M}^{(0)} \equiv (2\,e^2/h)\, C_0$ which 
depends on $v$. 
The normalized conductance decreases with $v$.
This also means that the effect of 
the interaction is enhanced when the level width of 
the resonant states is small.

\subsection{
Asymmetric connection $\,v_L \neq v_R$}
\label{subsec:asym_LR}

We next examine the case where the inversion symmetry is broken, 
$v_L^{\phantom{\dagger}} \neq v_R^{\phantom{\dagger}}$,
but the electron-hole symmetry is still preserved in the equilibrium state 
by the condition $\mu=0$ and $\epsilon_0=-U/2$. 
As one typical example, 
we take the mixing parameter to 
be $v_L^{\phantom{\dagger}}/t =0.8$ and $v_R^{\phantom{\dagger}}/t =0.6$ in 
this subsection.
In Fig.\ \ref{fig:g_N_asymm} (a) and (b),
the conductance is plotted as a function of the size $N$ 
taking $U/(2\pi t)$ to be (a) $0.2$ and (b) $0.4$, 
respectively. 
The result shows the even-odd oscillatory behavior 
as that in the case of $v_L^{\phantom{\dagger}}=v_R^{\phantom{\dagger}}$  
discussed in the above. 
However, in the case 
of $v_L^{\phantom{\dagger}} \neq v_R^{\phantom{\dagger}}$, 
the perfect transmission does not occur 
as it can be deduced from Eq.\ (\ref{eq:g_1_eff}), 
so that we have done the numerical calculation also for odd $N$. 
Although the Kondo resonance is still present 
at the Fermi level owing to the electron-hole symmetry, 
the conductance is reduced.
Nevertheless, the Kondo state contributes 
to the higher transmission for odd $N$.
In Fig.\ \ref{fig:g_N_asymm_U}, 
$g_{N}^{\phantom{\dagger}}$ is plotted as a function of $U$ 
for a number of $N$ ($=1,\,2,\,3,\,\ldots$,\,40).
For even $N$, qualitative features of the results are 
similar to those in the inversion symmetric case 
[See Fig.\ \ref{fig:g_2M_symm_U}].
On the other hand, for odd $N$, 
the features of the curves are somewhat different.
However, the results for large values of $U$ have some limitations,
because the parameter $\lambda$ defined by Eq.\ (\ref{eq:lambda})  
tends to an incorrect limit for large $U$ when the self-energy is estimated 
within a finite order in $U$, i.e., $\lambda$ tends to the ratio of 
the highest-order term of the numerator 
and that of the denominator of Eq.\ (\ref{eq:lambda_2}).
In the non-interacting case the conductance for even and odd $N$,  
$g_{2M}^{\phantom{\dagger}}$ and $g_{2M+1}^{\phantom{\dagger}}$, 
are independent of the size $M$, respectively, 
since  $\widetilde{v}_C = t$ and $\lambda=1$.  
For small $U$, the reduction of the conductance 
is proportional to $U^2$ and 
the curvature increases with the size $M$, 
except for the single impurity case $N=1$.  
In Fig. \ref{fig:c2_assym}, 
the coefficient $C_2$ defined by $g_{N}^{\phantom{\dagger}}= (2\,e^2/h) 
[ C_0 - C_2 \{U/(2\pi t)\}^2 + \cdots]$ is plotted as a function of $N$.
In the figure, the envelop for even $N$ is in upward from that for odd $N$.
The coefficient $C_2$ increases with the size $N$ showing 
an oscillatory behavior, and probably it diverges 
in the limit of large $N$.


\section{SUMMARY}
\label{sec:SUMMARY}

We have applied a quasi-particle description of a Fermi liquid 
to the transport through a small interacting system connected to reservoirs. 
With this approach, 
we have studied the properties of quasi-particles 
in an electron-hole symmetric case. 
In this case, effects of electron correlation 
come in the theory through the off-diagonal part of the self-energy, 
and the conductance can be written in a simplified form 
Eqs.\ (\ref{eq:g_1_eff}) or (\ref{eq:g_2_eff}) 
depending on whether the number of the interacting sites $N$
is even or odd. 
It is shown using Eq.\ (\ref{eq:g_1_eff})  that 
the perfect transmission occurs quite generally 
for odd $N$ if the system  has an inversion symmetry in addition 
to the electron-hole symmetry. 
We have applied the method to a small Hubbard chain of 
finite size $N$ with the second-order perturbation theory in $U$,
and examine the conductance as a function of $N$, $U$, 
and the coupling between the chain and the leads, i.e., 
$v_L^{\phantom{0}}$ and $v_R^{\phantom{0}}$. 
Effects of electron correlation are enhanced 
in weak coupling case $v_{L}^{\phantom{0}}, v_{R}^{\phantom{0}}<t$.
In this case, the conductance for even $N$ tends to zero for large $N$.
This behavior is qualitatively different from that  
in the special case $v_L^{\phantom{0}}=v_R^{\phantom{0}}=t$. 
This  seems to be caused by the difference in the structure of 
the resonant states. 
Quantitatively,  
the range of $U$ in which the second-order perturbation theory 
is able to provide reliable results tends to be narrow with increasing $N$,
and thus the higher order terms should be included
in order to refine the results obtained for large $U$ 
and to clarify the asymptotic behavior for large $N$.

Throughout this work, we have assumed that   
the interaction is switched on only in the sample region at the center.
When the interaction is switched on also in the two leads, 
the description in terms of the quasi-particles is still possible 
if the ground state is changed continuously against this interaction. 
But some modifications are necessary. 
The interaction in the leads will cause 
the renormalization of the incoming and outgoing fields 
corresponding to initial and final states of the scattering matrix, 
i.e., the renormalization of the external lines. 
Furthermore, the quasi-particle description can be 
extended to finite temperatures by taking into account 
the residual interaction among the quasi-particles 
or the contributions of vertex corrections.
Especially, an overall picture of the even-odd property 
described in the last part of Sec.\ \ref{sec:MODEL}  
will be confirmed microscopically based on 
the finite temperature theory.

\acknowledgments 
We would like to thank  
H. Fukuyama, H. Ishii, W. Izumida, N. Nagaosa, S. Nonoyama, 
and O. Sakai for valuable discussions. 
This work is partially supported by the Grand-in-Aid for Scientific Research 
from the Ministry of Education, Science and Culture, Japan.


\appendix

\section{$\mbox{\protect \boldmath ${\cal K}$}$ matrix
in the electron-hole symmetric case}
\label{sec:symmetric_app}

We summarize here properties 
of the renormalized hopping matrix $\mbox{\boldmath ${\cal K}$}$, 
and provide the derivation of Eq.\ (\ref{eq:K_relation}).
In the electron-hole symmetric case, 
$\mbox{\boldmath ${\cal K}$}$ has a checkered structure 
as shown in Eqs.\ (\ref{eq:K5}) and (\ref{eq:K6}).
Thus, $\det \mbox{\boldmath ${\cal K}$} =0$ for odd $N$.
For even $N$ ($=2M$), the determinant can be divided into two parts; 
\begin{equation}
\left | \, 
\matrix {  
 0  & \widetilde{t}_{12}^C & 0 & \widetilde{t}_{14}^C & \cdots \cr
 \widetilde{t}_{21}^C & 0  & \widetilde{t}_{23}^C      & 0 &     \cdots \cr
 0  & \widetilde{t}_{32}^C  & 0 & \widetilde{t}_{34}^C & \cdots \cr
 \widetilde{t}_{41}^C & 0  & \widetilde{t}_{43}^C      & 0 &  \cdots \cr
 \vdots             & \vdots  & \vdots & \vdots & \ddots \cr
        }
\, \right |  
\ = \  
(-1)^{M} 
\det \mbox{\boldmath $Q$} \, 
\det \mbox{\boldmath $Q$}' 
\;.
\label{eq:K_Factor} 
\end{equation}
Here $\mbox{\boldmath $Q$}$  and  $\mbox{\boldmath $Q$}'$ 
are $M \times M$ matrices defined by 
\begin{equation}
\mbox{\boldmath $Q$} = 
\left [ \, 
\matrix { \widetilde{t}_{12}^C  & \widetilde{t}_{14}^C  &  \cdots  \cr
          \widetilde{t}_{32}^C  & \widetilde{t}_{34}^C  &  \cdots  \cr
          \vdots & \vdots &  \ddots  \cr
       }
\, \right ] 
\;,  \qquad 
\mbox{\boldmath $Q$}' =  
\left [ \, 
\matrix { \widetilde{t}_{21}^C  & \widetilde{t}_{23}^C  &  \cdots  \cr
          \widetilde{t}_{41}^C  & \widetilde{t}_{43}^C  &  \cdots  \cr
          \vdots & \vdots &  \ddots  \cr
       }
\, \right ]  \;.
\label{eq:Q}
\end{equation}
Since the renormalized parameter is real 
and $\widetilde{t}_{ij}^C=\widetilde{t}_{ji}^C$ owing to 
the time-reversal symmetry, 
$\det \mbox{\boldmath $Q$} = \det \mbox{\boldmath $Q$}'$.  
Consequently, for even $N$ ($=2M$),  
the determinant of the matrices $\mbox{\boldmath ${\cal K}$}$,  
$\mbox{\boldmath ${\cal K}$}_{11}^{NN}$, 
and $\mbox{\boldmath ${\cal K}$}_{N1}$ can be factorized as 
\begin{eqnarray}
\det \mbox{\boldmath ${\cal K}$} &=& (-1)^M \, 
\left\{ \det \mbox{\boldmath $Q$} \right\}^2\;,  
\label{eq:K_even}
\\
\det \mbox{\boldmath ${\cal K}$}_{11}^{NN} &=& (-1)^{M-1} \, 
\left\{ \det \mbox{\boldmath $Q$}_{11}^{NN} \right\}^2\;,  
\\
\det \mbox{\boldmath ${\cal K}$}_{N1} &=& 
\det \mbox{\boldmath $Q$} \, 
\det \mbox{\boldmath $Q$}_{11}^{NN} 
\label{eq:K_N1_even}
\;. 
\end{eqnarray}
Here $\mbox{\boldmath $Q$}_{11}^{NN}$ is 
a $(M-1) \times (M-1)$ matrix extracted from 
$\mbox{\boldmath ${\cal K}$}_{11}^{NN}$ in the similar way 
as it was done for extracting $\mbox{\boldmath $Q$}$ 
from $\mbox{\boldmath ${\cal K}$}$. 
Using Eqs.\ (\ref{eq:K_even})--(\ref{eq:K_N1_even}), 
we obtain Eq.\ (\ref{eq:K_relation}) for even $N$, i.e., 
$(\det \mbox{\boldmath ${\cal K}$}_{N1})^2 =  
-\det \mbox{\boldmath ${\cal K}$}
\, \det \mbox{\boldmath ${\cal K}$}_{11}^{NN}$.
Furthermore, $\widetilde{v}_{C}$ defined by Eq.\ (\ref{eq:v_eff}) is 
simplified as 
\begin{equation}
\widetilde{v}_{C} 
\ = \  \left|\, {\det \mbox{\boldmath $Q$} \over
              \det \mbox{\boldmath $Q$}_{11}^{NN} } \,\right|
              \;.
\label{eq:v_eff_2}
\end{equation}

For odd $N$ ($=2M+1$), 
the $2M \times 2M$ matrices 
$\mbox{\boldmath ${\cal K}$}_{11}$ and 
$\mbox{\boldmath ${\cal K}$}_{NN}$ 
have the similar properties described in the above. 
Consequently, the determinants can be factorized as
\begin{eqnarray}
\det \mbox{\boldmath ${\cal K}$}_{11} &=& (-1)^M \, 
\left\{ \det \mbox{\boldmath $Q$}_{11} \right\}^2\;,  
\label{eq:K_odd}
\\
\det \mbox{\boldmath ${\cal K}$}_{NN} &=& (-1)^M \, 
\left\{ \det \mbox{\boldmath $Q$}_{NN} \right\}^2\;,  
\\
\det \mbox{\boldmath ${\cal K}$}_{N1} &=& 
\det \mbox{\boldmath $Q$}_{11} \, 
\det \mbox{\boldmath $Q$}_{NN} 
\label{eq:K_N1_odd}
\;. 
\end{eqnarray}
Here $\mbox{\boldmath $Q$}_{11}$ 
and $\mbox{\boldmath $Q$}_{NN}$ are $M \times M$ matrices 
extracted from $\mbox{\boldmath ${\cal K}$}_{11}$ 
and $\mbox{\boldmath ${\cal K}$}_{NN}$, respectively, 
as it was done in the above. 
Using Eq.\ (\ref{eq:K_odd})--(\ref{eq:K_N1_odd}), 
we obtain Eq.\ (\ref{eq:K_relation}) for odd $N$, i.e., 
$(\det \mbox{\boldmath ${\cal K}$}_{N1})^2 =  
\det \mbox{\boldmath ${\cal K}$}_{11}
\, \det \mbox{\boldmath ${\cal K}$}_{NN}$.  
Also, $\lambda$ defined by Eq.\ (\ref{eq:lambda}) is simplified as
\begin{equation}
\lambda = 
\left| 
{ \det \mbox{\boldmath $Q$}_{11} \over
  \det \mbox{\boldmath $Q$}_{NN} } 
\right|
\;.
\label{eq:lambda_2}
\end{equation}

\section{reflection coefficient in a special case}
\label{sec:reflection}

We show here another approach to clarify 
the asymptotic behavior of the coefficient $C_4$ 
introduced in Sec.\ \ref{sec:2nd_order} 
in the case of $v_L^{\phantom{\dagger}} = v_R^{\phantom{\dagger}} =t$. 
In this case, 
the Dyson equation Eq.\ (\ref{eq:Dyson}) 
is written in terms of the scattering matrix ${\cal T}_{ll'}$ as,  
\begin{equation} 
  G_{jj'}  \  
= \  G^{(0)}_{jj'}
    + \sum_{l,l'=1}^{N}\,G^{(0)}_{jl}
   \ {\cal T}_{ll'}
   \ G^{(0)}_{l'j'} \;.
  \label{eq:Dyson_T}
\end{equation} 
The transmission and reflection coefficients are also
written in terms of  ${\cal T}_{ll'}$ 
as   $\widetilde{t}(0) = 1 - i\,{\cal T}_{k_F k_F}(i0^+)/v_F$ 
and  $\widetilde{r}(0) = -i \, {\cal T}_{-k_F k_F}(i0^+)/v_F$, 
\cite{ao5,ao6} where $v_F=2t$,  $k_F=\pi/2$, and 
\begin{equation}
{\cal T}_{kk'} = \displaystyle \sum_{l,l'=1}^{N}\, 
e^{-i\,k l}\,{\cal T}_{ll'}\,e^{i\, k' l' }
\;.
\label{eq:T_kk'}
\end{equation}
Since the unperturbed system has  
the translational invariance in the present case,
the lowest-order scattering matrix 
is given by ${\cal T}^{(2)}_{ll'} = \Sigma^{(2)}_{ll'}$.
Thus, using a wavenumber representation of $\Sigma^{(2)}_{ll'}$,\cite{ao7}
the lattice sum Eq.\ (\ref{eq:T_kk'}) can be done explicitly, 
and the reflection coefficient is expressed as
\begin{eqnarray}
\widetilde{r}^{(2)}_N
\ &=& \ -{i\over v_F}\, e^{i\,k_F (N+1)}
\,U^2 
\int {dk_3 dk_2 dk_1 \over(2\pi)^3}  \,  
\ {          \cos k_F N - \cos [(k_1 + k_2 -k_3)N]    
           \over    \cos k_F - \cos (k_1 + k_2 -k_3) 
         }
 \nonumber \\
& &  \   
\times 
 \,\mbox{P}\, { f_{k_3}(1-f_{k_2})(1-f_{k_1}) + (1-f_{k_3})f_{k_2}f_{k_1}  
                \over
                 \xi_{k_3} -\xi_{k_2} -\xi_{k_1} 
                   }  
\;.
\label{eq:reflection}
\end{eqnarray}
Here 
$\xi_k = -2t \cos k$, $f_k = [e^{\beta \xi_k}+1]^{-1}$, 
and $\mbox{P}$ denotes the Cauchy principal value. 
It can be confirmed from Eq.\ (\ref{eq:reflection}) 
that $\widetilde{r}^{(2)}_N =0$ for odd $\,N$. 
For large even $N$, 
the contribution of the fast varying $\cos [(k_1 + k_2 -k_3)N]$ part 
becomes small, and the dominant contribution comes form 
the $\cos k_F N$ part.
Thus, for large even $N$ ($=2M$), 
the reflection coefficient is written in the form 
$\displaystyle \lim_{M\to \infty}\, \widetilde{r}^{(2)}_{2M} 
 =  \sqrt{C_4^{\infty}} \  \{U/(2\pi t)\}^2$ with 
\begin{equation}
\sqrt{C_4^{\infty}}   
\ \equiv {1\over 4\pi}\, \mbox{P} \int_{-\pi/2}^{\pi/2} 
{
    dk_3 dk_2 dk_1  \over 
    \left[\, \cos k_1 \,  + \cos k_2 \, + \cos k_3\right]\,
      \cos \left( k_1  + k_2  +  k_3 \right) 
}
\;.
\label{eq:A_lattice} 
\end{equation}
The coefficient $\sqrt{C_4^{\infty}}$ is finite 
and estimated numerically as $\sqrt{C_4^{\infty}} \simeq 0.729$. 
Thus, in the limit of large even $N$, 
the dc conductance 
in the case of $v_L^{\phantom{\dagger}} = v_R^{\phantom{\dagger}} =t$ 
is written in the form  
$\displaystyle \lim_{M\to \infty}\, g_{2M} 
= (2e^2 / h) \, \left[\, 1 - C_4^{\infty}\, \{U/(2\pi t)\}^4 \, 
                               + \, \cdots \, \right]$.

\begin{figure}
\caption{ Schematic picture of the system} 
\label{fig:single}
\end{figure}

\begin{figure}
\caption{ Schematic picture of the model: 
($\bullet$) interacting region, ($\circ$) ideal leads.} 
\label{fig:lattice}
\end{figure}

\begin{figure}
\caption{ Second-order self-energy $\Sigma^{(2)}_{ij}(i\omega)$.}
\label{fig:diagram}
\end{figure}

\begin{figure}
\caption{Conductance $g_{N}^{\protect\phantom{\dagger}}$ as 
a function of the size $N$ in the inversion symmetric case. 
Here $v/t = 0.8$ and $U/(2 \pi t) = 0.4$. 
 }
\label{fig:g_N_symm}
\end{figure}

\begin{figure}
\caption{Conductance for even $N$ ($=2M$) as a function of the size $M$. 
Here $v/t=0.8$, and the repulsion $U/(2 \pi t)$ is taken to be 
($\bullet$) $0.0$, ($\blacksquare$) $0.2$, ($\blacklozenge$) $0.4$,
($\blacktriangle$) $0.6$, and ($\bigstar$) $0.8$. 
}
\label{fig:g_2M_symm_several_U}
\end{figure}

\begin{figure}
\caption{
Conductance vs $U$, for even $N$ ($=2,\,4,\,6,\,\ldots$).
Here $v/t=0.8$ for dashed lines, and $v/t=1.0$ for solid lines. 
Totally, $20$ dashed and $30$ solid lines are plotted.
}
\label{fig:g_2M_symm_U}
\end{figure}

\begin{figure}
\caption{The ratio $C_2/C_0$ is plotted 
for several $v/t$ (=$0.1,\, 0.2, \, \ldots,\, 0.9$) 
as a function of the size $M$.
Here $g_{2M}^{\protect \phantom{\dagger}}= (2\,e^2/h)  
[\, C_0 - C_2 \{ U/(2\pi t) \}^2 + \cdots \,]$. 
}
\label{fig:c2_c0}
\end{figure}

\begin{figure}
\caption{ 
The coefficient $C_4$ in the case of $v/t= 1.0$ is plotted  
as a function of the size $M$.
Here $g_{2M}^{\protect \phantom{\dagger}}= (2\,e^2/h)  
[\, 1 - C_4 \{ U/(2\pi t) \}^2 + \cdots \,]$. 
}
\label{fig:c4}
\end{figure}

\begin{figure}
\caption{ Conductance vs $U$, for $N=4$ (dashed lines) 
and $N=40$ (solid lines).
Here $v/t$ is taken to be $0.4,\,0.5,\, \ldots ,\, 0.9$, and  $1.0$.
}
\label{fig:g_2M_symm_U2}
\end{figure}

\begin{figure}
\caption{
Conductance for even $N$ ($=2M$) as a function of the size $M$. 
Here $U/(2 \pi t) = 0.3$, and 
the conductance is normalized by
the value for $U=0$, i.e., $g_{2M}^{(0)} \equiv (2e^2/h)\, C_0$. 
The mixing matrix element $v/t$ is taken to be  
($\blacksquare$) $0.4$, ($\bullet$) $0.5$
($\bigstar$) $0.6$, 
($\blacktriangle$) $0.7$,
($\blacklozenge$) $0.8$,
($\blacksquare$) $0.9$, and ($\bullet$) $1.0$.
}
\label{fig:g_2M_symm_several_V}
\end{figure}

\begin{figure}
\caption{
Conductance $g_{N}^{\protect\phantom{\dagger}}$ as 
a function of the size $N$ in the inversion asymmetric case. 
Here $v_L^{\protect\phantom{\dagger}}/t=0.8$, 
$v_R^{\protect\phantom{\dagger}}/t =0.6$, 
and $U/(2 \pi t)$ is taken to be (a) $0.2$ and (b) $0.4$.
 }
\label{fig:g_N_asymm}
\end{figure}

\begin{figure}
\caption{
Conductance vs $U$,    
for a number of $N$ ($=1,\,2,\,3,\,\ldots$,\,40).
Here $v_L^{\protect\phantom{\dagger}}/t=0.8$ 
and $v_R^{\protect\phantom{\dagger}}/t =0.6$.
}
\label{fig:g_N_asymm_U}
\end{figure}

\begin{figure}
\caption{
The coefficient $C_2$ as a function of the size $N$.
Here $v_L^{\protect\phantom{\dagger}}/t=0.8$, 
$v_R^{\protect\phantom{\dagger}}/t =0.6$, and 
 $g_{N}^{\protect \phantom{\dagger}}= (2\,e^2/h)  
[\, C_0 - C_2 \{ U/(2\pi t) \}^2 + \cdots \,]$. 
}
\label{fig:c2_assym}
\end{figure}

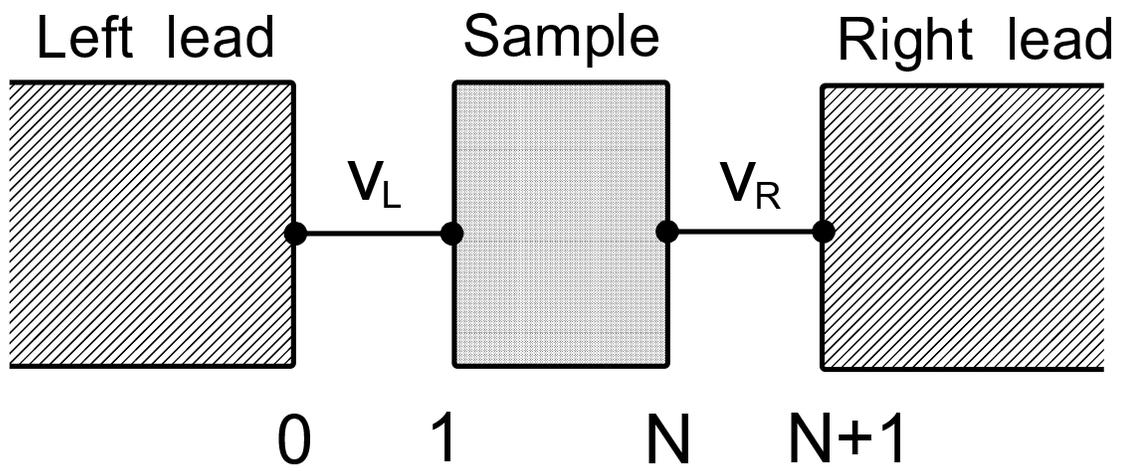
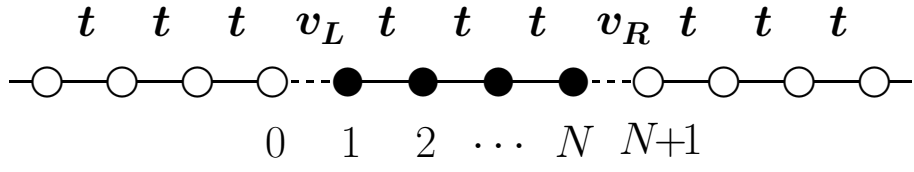
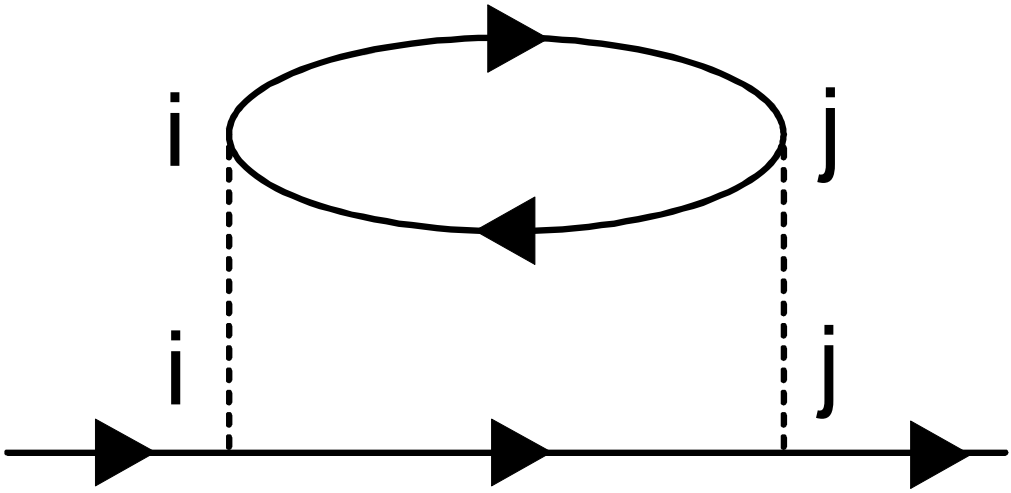
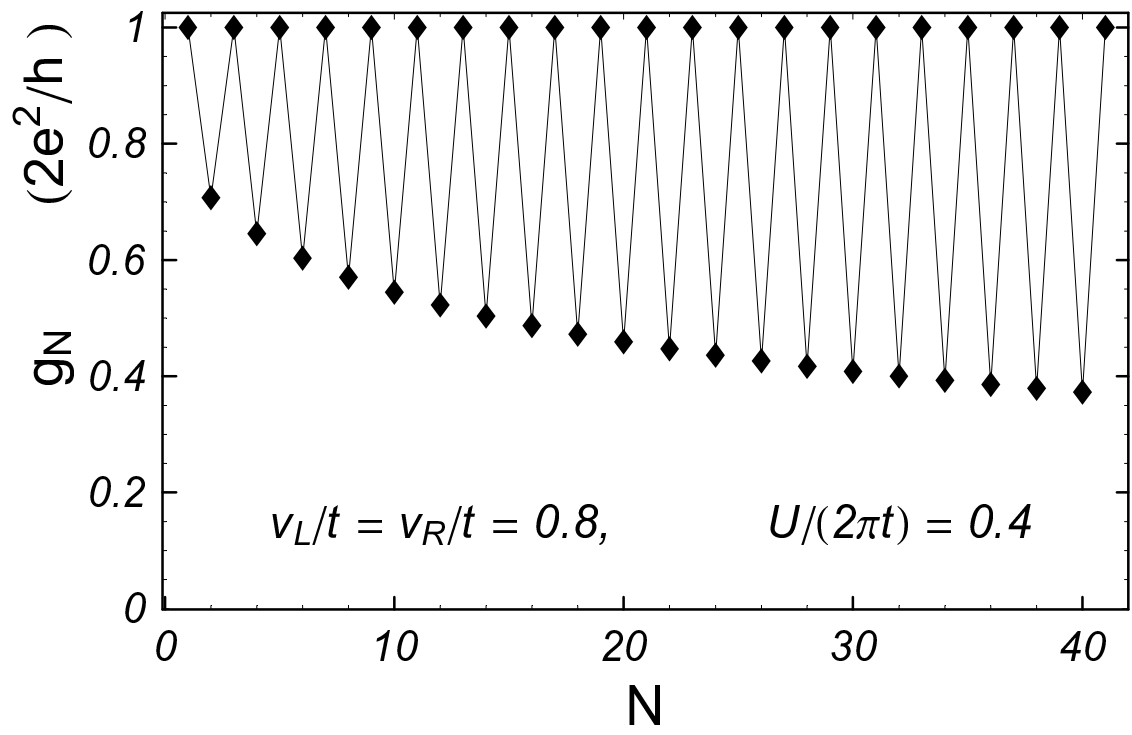
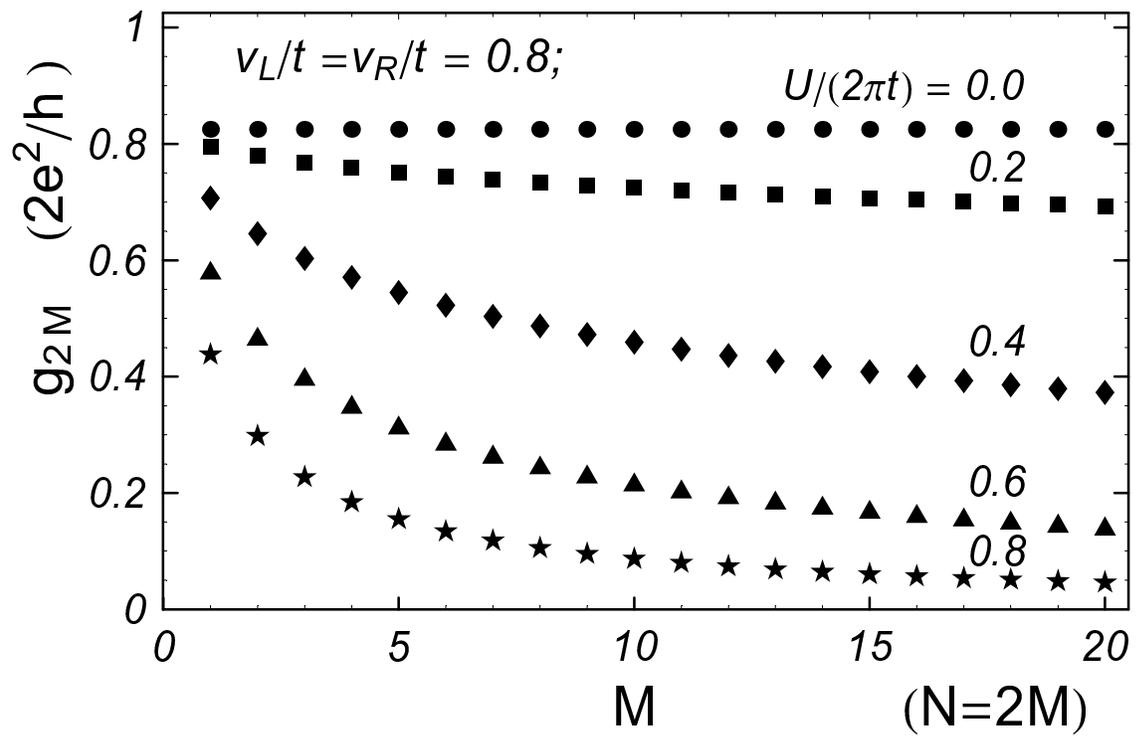
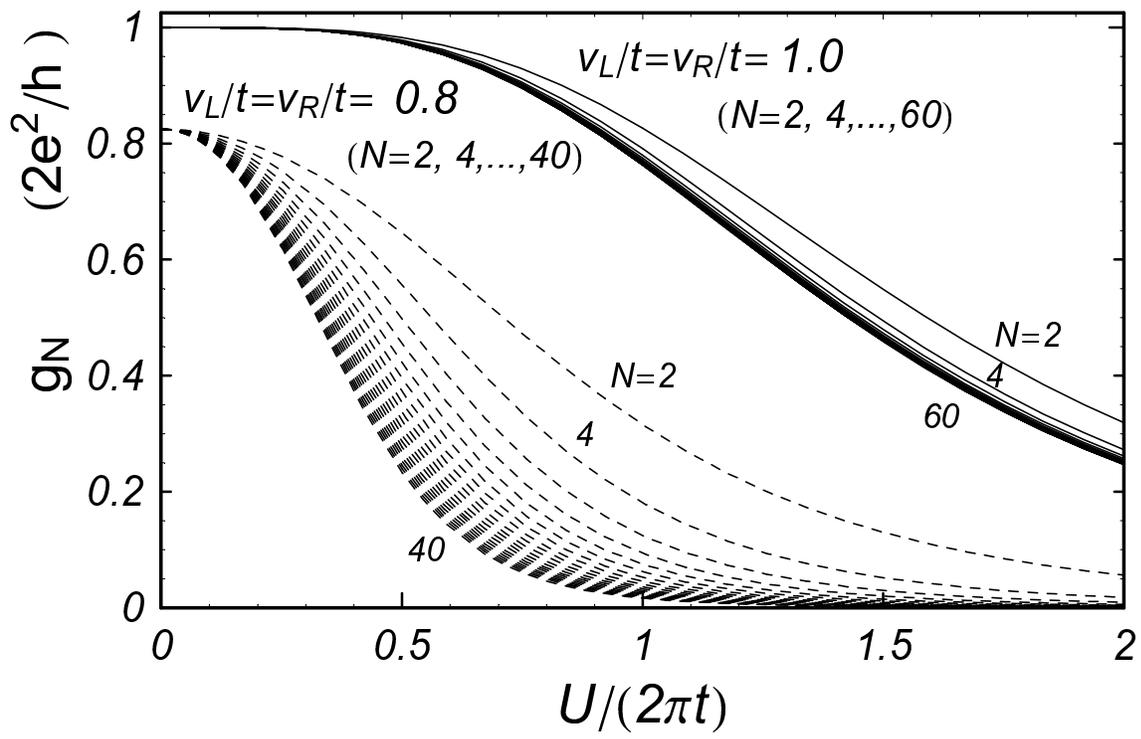
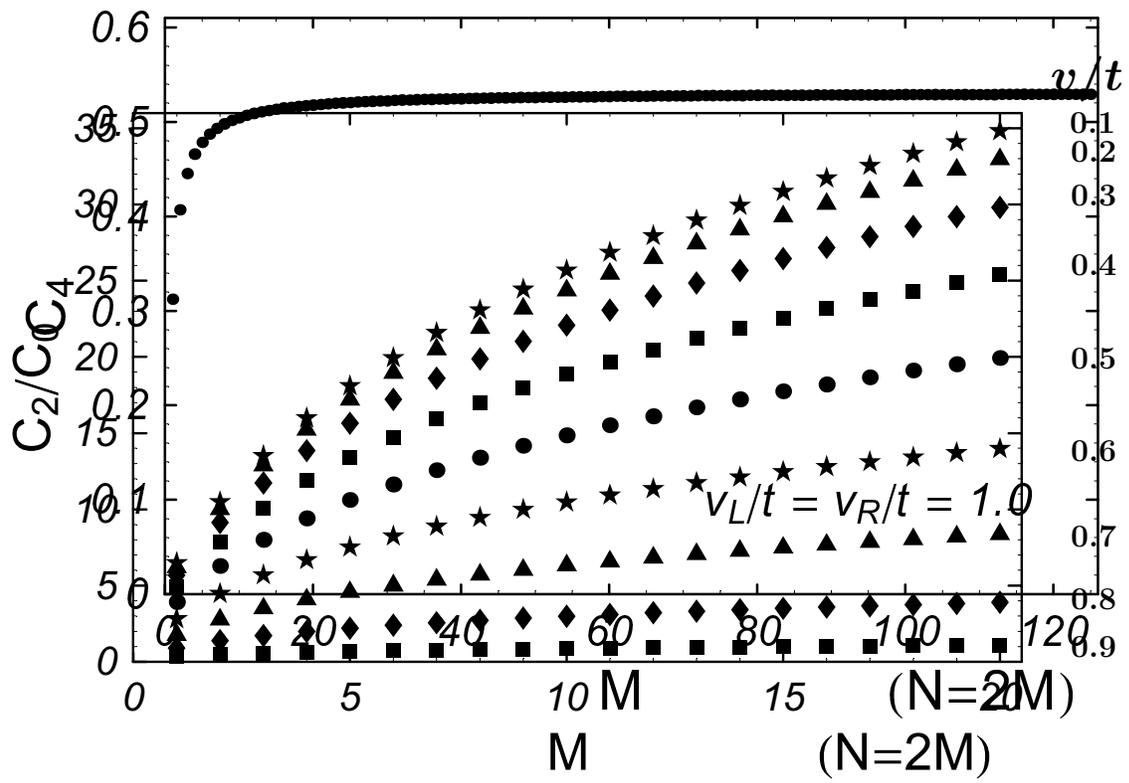
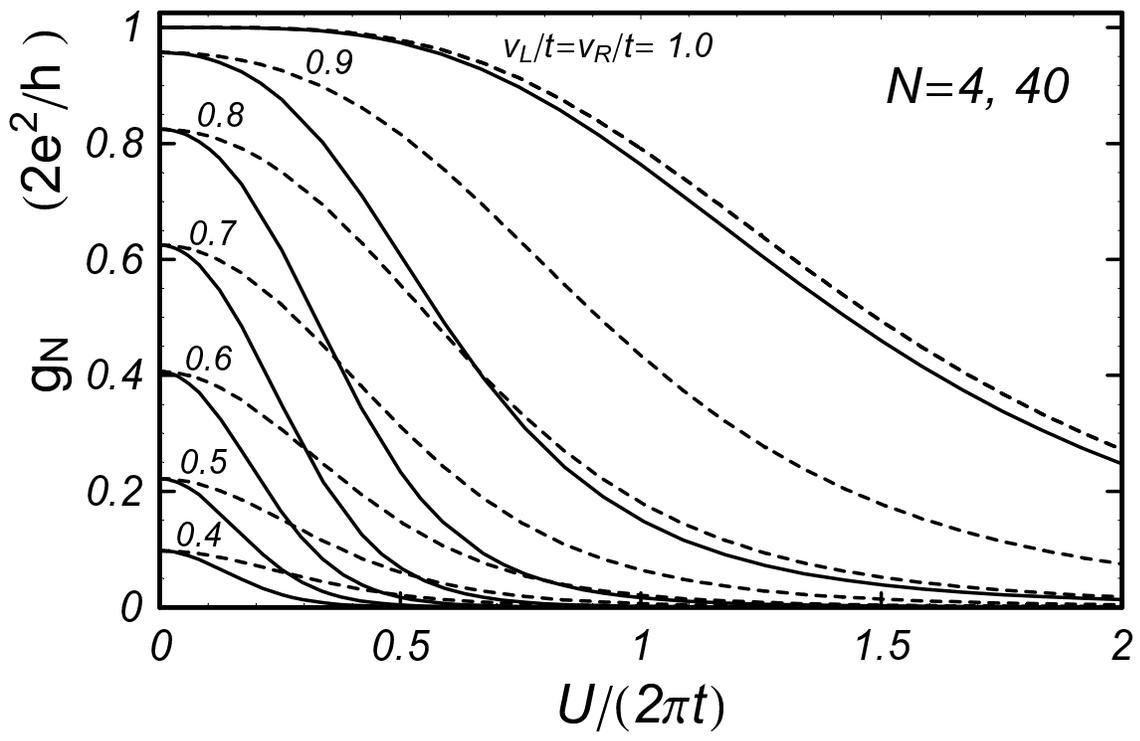
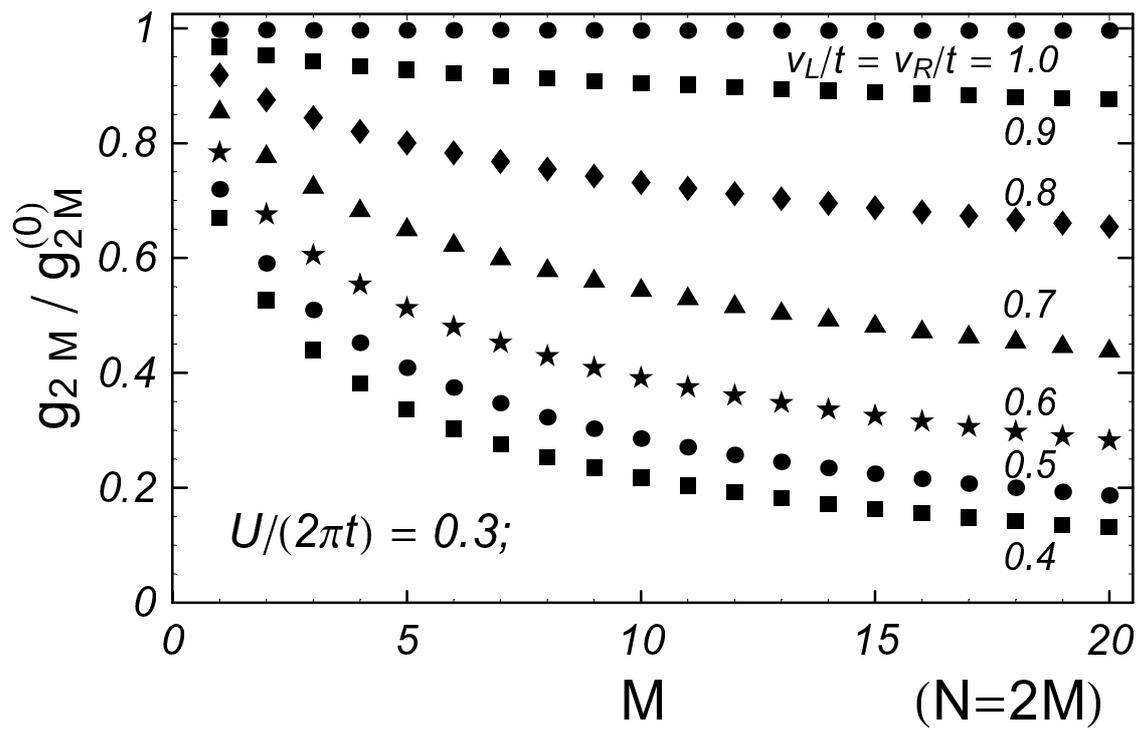
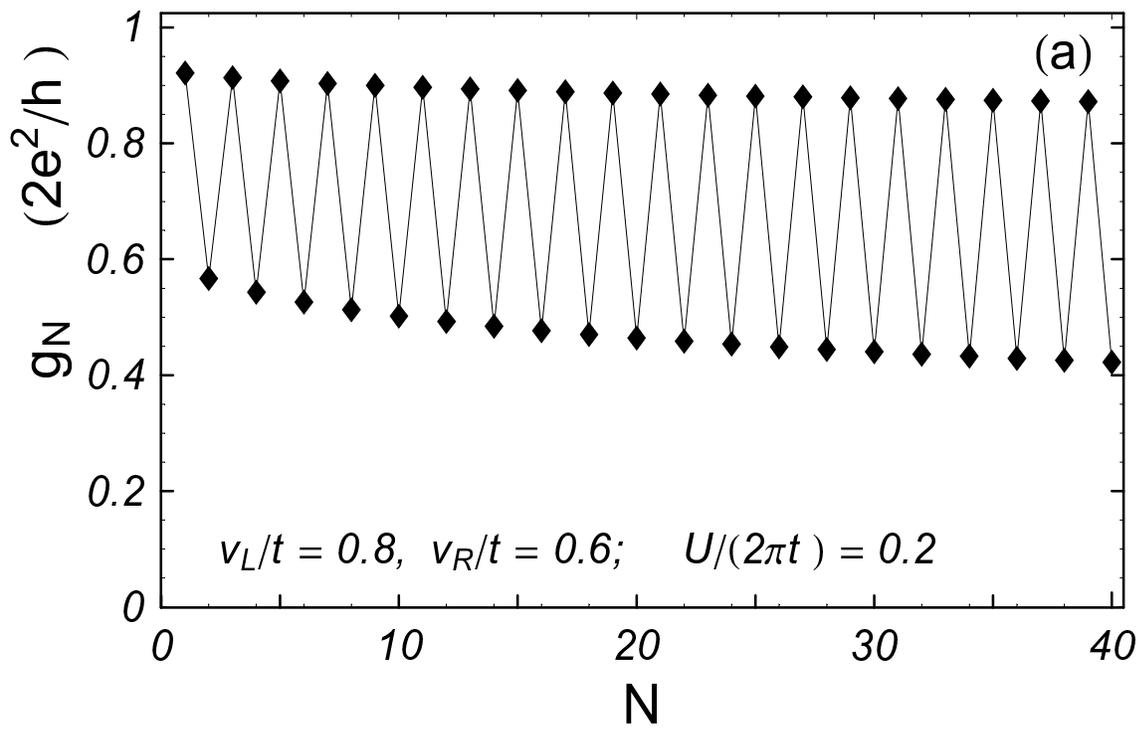
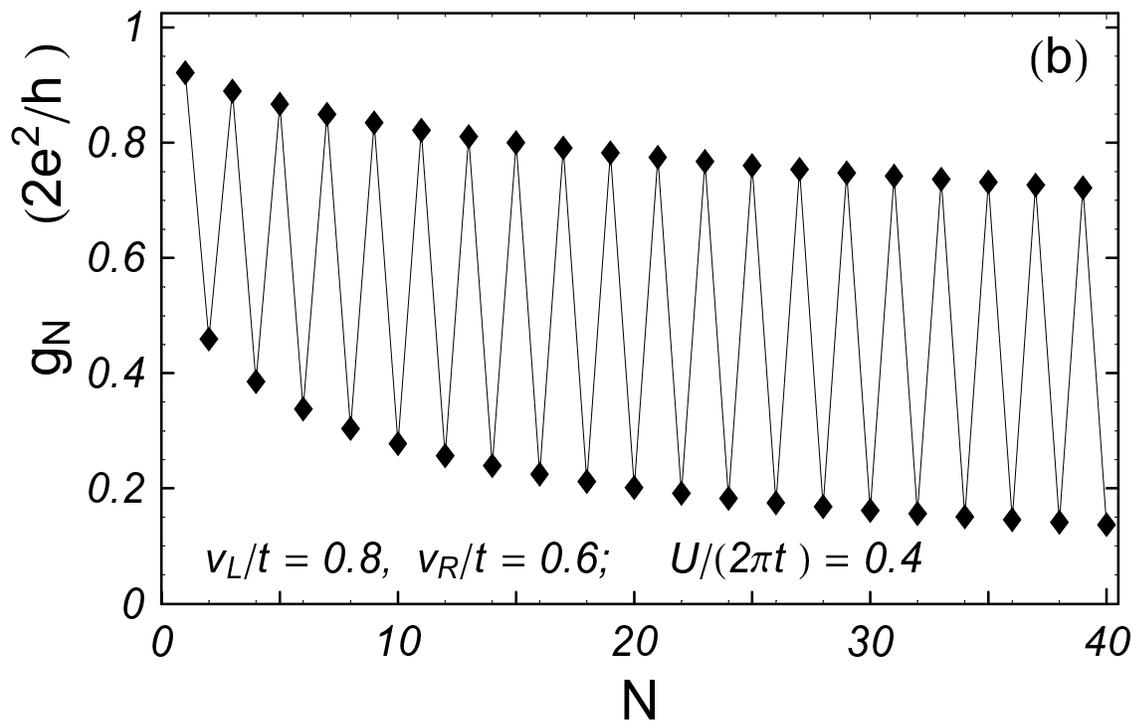
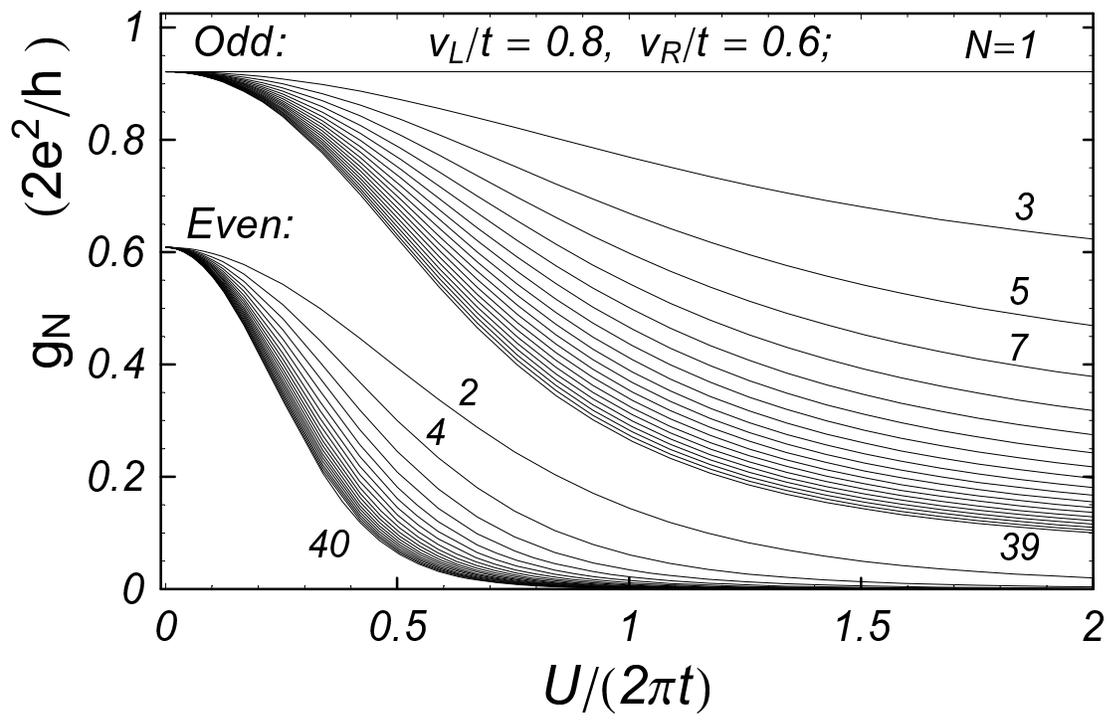
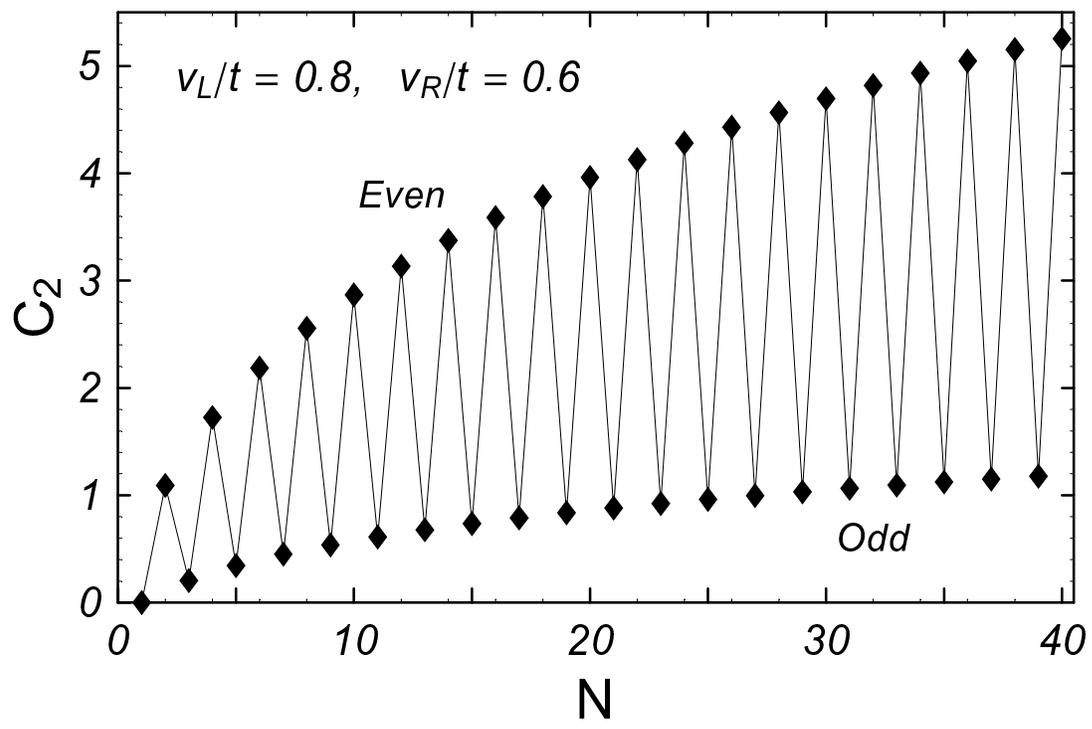
\end{document}